# Throughput in Asynchronous Networks

Paul Bunn     Rafail Ostrovsky

November 20, 2018

**Abstract**

We introduce a new, "worst-case" model for an asynchronous communication network and investigate the simplest (yet central) task in this model, namely the feasibility of end-to-end routing. Motivated by the question of how successful a protocol can hope to perform in a network whose reliability is guaranteed by as few assumptions as possible, we combine the main "unreliability" features encountered in network models in the literature, allowing our model to exhibit all of these characteristics *simultaneously*. In particular, our model captures networks that exhibit the following properties:

- On-line
- Dynamic Topology
- Distributed/Local Control
- Asynchronous Communication
- (Polynomially) Bounded Memory
- No Minimal Connectivity Assumptions

In the confines of this network, we evaluate throughput performance and prove matching upper and lower bounds. In particular, using competitive analysis (perhaps somewhat surprisingly) we prove that the optimal competitive ratio of any on-line protocol is $1/n$ (where $n$ is the number of nodes in the network), and then we describe a specific protocol and prove that it is $n$-competitive.

The model we describe in the paper and for which we achieve the above matching upper and lower bounds for throughput represents the "worst-case" network, in that it makes no reliability assumptions. In many practical applications, the optimal competitive ratio of $1/n$ may be unacceptable, and consequently stronger assumptions must be imposed on the network to improve performance. However, we believe that a fundamental starting point to understanding which assumptions are necessary to impose on a network model, given some desired throughput performance, is to understand what is achievable in the *worst case* for the simplest task (namely end-to-end routing). Additionally, our work may also serve as a framework to study additional questions (beyond end-to-end communication) within this "worst-case" case network setting.

# 1 Introduction

With the immense range of applications and the multitude of networks encountered in practice, there has been an enormous effort to study routing in various settings. In terms of understanding what routing standards are possible, researchers attack the problem from two ends: understanding what is optimal, and developing and evaluating protocols. Not surprisingly, provable results at both ends rely heavily on the model chosen to capture the features of the network.

Typically, networks are modelled as a graph with vertices representing nodes (processors, routers, etc.) and edges representing the connections between them. Beyond this basic structure, additional assumptions and restrictions are then made in attempt to capture various features that real-world networks may display. In particular, a network model must specify a choice between each of the following characteristics:

|  | Weak Assumption | Strong Assumption |
|---|---|---|
| View of Network: | On-line | Off-line |
| Edge-Reliability: | Dynamic Topology | Fixed Topology |
| Control Mechanism: | Distributed/Local Control | Centralized/Global Control |
| Timing/Coordination: | Asynchronous Communication | Synchronous Communication |
| Resources: | Bounded Memory | Unlimited Memory |
| "Liveness": | No Connectivity Assumptions | Minimal Connectivity Guarantees |

Notice that in each option above there is an inherent trade-off between generality/applicability of the model verses optimal performance within the model. For instance, a protocol that assumes a fixed network topology will likely out-perform a protocol designed for a dynamic topology setting, but the former protocol may not work in networks subject to edge-failures. From both a theoretical and a practical standpoint, it is important to understand how each (combination) of the above factors affects routing performance.

In this paper, we study the feasibility of routing in an "unrestricted" network model: one that simultaneously considers *all* of the more general features from the above list. Admittedly, in this "worst-case" model it is unlikely that any protocol will perform well, and one (or more) stronger assumptions must be made to achieve a reasonable level of performance. However, understanding behavior in the worst case, even with respect to the most basic task of end-to-end communication, is important to determine how much (if any) the addition of each strong assumption improves optimal protocol performance.

## 1.1 Previous Work

To date, all network models have made at least one of the (and more commonly multiple) strong assumptions listed above. The amount of research regarding network routing and analysis of routing protocols is extensive, and as such we include only a sketch of the most related works, indicating how their models differ from ours and providing references that offer more detailed descriptions.

<u>Competitive Analysis</u>    Competitive Analysis was first introduced by Sleator and Tarjan [34] as a mechanism to analyze a specific protocol in terms of how vulnerable the protocol is to being out-performed by alternative protocols. Notice that with respect to the four choices for modeling a network, three choices can be represented by *imperfect information* available to any *on-line* protocol that operates within the network. In particular, in a dynamic network, a protocol does not know the future topology of the network; in a distributed network, nodes do not know the state of the other nodes and/or the links of the network; and in an asynchronous network, a protocol does not know how long packets will take to traverse a link. Recall that a given protocol has *competitive ratio* $1/\lambda$ (or is $\lambda$-*competitive*) if an ideal off-line protocol has advantage over the given protocol by at most a factor of $\lambda$. Competitive analysis may be applied to any of the possible network models, but our work is the first to utilize this analysis in the unrestricted network model considered in this paper. For a thorough description of competitive analysis, see [16].

End-to-End Communication    In this paper, we focus on the task of performing throughput-efficient End-to-End Communication. Although there is an enormous amount of work in developing and analyzing protocols for end-to-end-communication, analysis in terms of throughput performance has been restricted to *synchronous* networks.[1]

Max-Flow and Multi-Commodity Flow    The Max-flow problem assumes networks that are: fixed topology, global-control, and synchronous. This is far more restrictive a network model than the one we consider in this paper.

The multi-commodity flow model (see for example Leighton et al. [30]), is a generalization of the max-flow model, in which the network being considered may be dynamic and distributed. However, even if the network is dynamic, at every round the topology of the network is assumed to be adequate to meet the demands of all commodities.[2]

In the work of Awerbuch and Leighton [13], competitive analysis is utilized to demonstrate a protocol for multi-commodity flow which is $(1 + \epsilon)$-competitive.[3] In particular, they guarantee that if the ideal off-line protocol can meet $z(1 + 3\epsilon)$ percent of the demand of every commodity at all times, then their protocol will meet $z(1 + \epsilon)$ percent of the demand of every commodity. Since a linear competitive ratio is the best *any* protocol can hope to achieve, their result is very close to optimal (in this network model).

Admission Control and Route Selection    For an extensive discussion about research in the area of Admission Control and Route Selection, see [31] and references therein. The admission control/route selection model differs from the multi-commodity flow model in that the goal of a protocol is not to meet the demand of all ordered pairs of nodes $(s,t)$, but rather the protocol must decide which requests it can/should honor, and then designate a path for honored requests.

There are numerous variants to the basic admission control/route selection problem, including Probabilistic Analysis (requests are described via distributions) [28] and [29], Routing with Edge-Congestion [21], Multi-cast Routing [9], [15], and [25], Unsplittable Flow [20], Load Control Problems [5], Non-Blocking Networks [7], and Ad Hoc Networks [12] and [32]. Competitive analysis was first utilized in the admission control/route selection model to analyze throughput by Garay and Gopal [23] and Garay et al. [24] for sparse networks, and later for more general networks by a series of authors, including [10], [8], and [26]. Competitive analysis has also been used to measure theoretical bounds on throughput performance for multi-casting in synchronous networks in [11].

In the Asynchronous Transfer Model (ATM), the (commodity, demand) requests from pairs of nodes $(s,t)$ are not available at the outset of the protocol, and furthermore these requests are *temporary*, with a given *duration* of time for which $s$ wishes to transfer the commodity to $t$. We emphasize that the definition of *asynchronicity* in ATM is different than the one considered in this paper. In particular, "asynchronicity" in ATM literature is meant to emphasize the fact that the requests are *not* known ahead of time, and thus protocols face the added challenge of handling new requests adaptively. As in the admission control/route selection model, the goal of a routing protocol in the ATM is to decide which requests to honor (and how to route the honored requests) based on the current status of the network, such that all previously honored requests are not deleted. In the work of Awerbuch, Azar, and Plotkin [10], competitive analysis is utilized to demonstrate a protocol that is $\log(nT)$-competitive[4] in terms of throughput, where $n$ is the number of nodes in the network and $T$ is an upper-bound on the call duration of any request.

---

[1] There are several message-driven protocols designed to work in asynchronous networks, for example the Slide protocol introduced by Afek and Gafni [2] (and further developed by [14], [1], and [27]). While these protocols work well in practice and have proven results in synchronous networks, to date there has been no rigorous throughput-analysis of these protocols in asynchronous networks.

[2] Alternatively, the *concurrent flow* problem considers networks whose topology may not be adequate to simultaneously meet the demands of all quantities. In this case the problem seeks to maximize the percentage $z$ such that at least $z$ percent of the demand of every commodity is met.

[3] Since the model assumes a dynamic and distributed network, the perfect information available to the off-line protocol includes knowledge of the future topology (which links will be available each round) as well as a global-view of the network.

[4] The perfect information available to the off-line protocol includes knowledge of all future requests.

The admission control/route selection model assumes a fixed topology, synchronous network (and is also often considered to be global-control). In addition, when a request $(s,t)$ is honored, often it is demanded that a routing path must be specified upon acceptance of the request, and this path is not permitted to change in an adaptive manner (e.g. a different path may not be utilized in order to optimize throughput as a result of a future requests).

**Queuing Theory**  A model related to the ATM is the Queuing Theory (QT) model (see e.g. [17] and [6]). In this model (as in the ATM), the pattern of requests (chosen from some known distribution) is not known ahead of time, but unlike ATM, *all* requests must be honored. Queuing Theory asks the question: given a distribution from which the requests will be chosen, how much memory does a given protocol require of each processor in order to guarantee all requests can be honored?

In alternate instantiations, the queuing theory model may either allow adaptive routing, or require that paths be fixed upon receipt of a request; also, there are works considering both global-control and distributed networks. Most analysis in the QT model has considered fixed topology networks, and in all models, the network is assumed to be synchronous.

**Adversarial Queuing Theory**  Adversarial Queuing Theory (AQT) is similar to QT, except the pattern of requests is controlled by an adversary who may wish to disrupt a given protocol as much as possible, and then competitive analysis is used to analyze performance. There has been a lot of work in the AQT model (e.g. [18] and [3]) that considers networks allowing dynamic topology (although the network is always assumed to be synchronous).

**Competitive Analysis of Distributed Algorithms**  So far, none of the above models have considered asynchronous networks. However, there has been a tremendous amount of effort to analyze distributed algorithms in asynchronous[5] shared memory computation, including the work of Ajtai et al. [4]. This line of work has a different flavor than the problem considered in the present paper due to the nature of the algorithm being analyzed (computation algorithm verses network routing protocol). In particular, network topology is not a consideration in this line of work.

## 1.2  Motivation

One of the primary goals of all routing protocols is to achieve high throughput between a Sender and a Receiver in a network, subject to resource constraints. As seen above, to date almost all of the research analyzing the throughput-performance of routing protocols has focused on synchronous networks, and the majority of works also assume fixed topology. However, in many practical settings, such as the Internet, the networks encountered are fully *asynchronous*, and network topology is typically unstable, with edges that may go up an down in an unpredictable manner. In the next few paragraphs, we discuss some of the challenges in designing and analyzing throughput performance of protocols in our "unrestricted" network model.

The metric in which we will wish to analyze and compare protocols is in the rate of packet delivery, or *throughput*. Intuitively, the throughput of a protocol measures the amount of information delivered as a function of time. However, capturing this intuition formally in the context of our network model encounters two difficulties. First, the *asynchronous* nature of the network makes it difficult to formalize the notion of "time." In a synchronous network, throughput is typically measured by first defining discrete "rounds," during which every connection can transfer one unit of information.[6] Throughput in the synchronous setting is then defined to measure the amount of information received as a function of the number of rounds that have passed. However, in an asynchronous network, there is no a priori notion of a round, as connections may be transferring information at different times and rates.[7] Therefore, capturing the intuition of what it means for a protocol to be efficient requires

---
[5]The terminology of "asynchronicity" here is consistent with the one defined in this paper, as opposed to the asynchronous transfer model (ATM) described above.

[6]Depending on the model, this unit may be the same for all connections, or it may vary for the different connections (in which case the terms "bandwidth" or "edge-capacity" are used). Also, the bandwidth for a single connection may be fixed for all time or may change each round.

[7]Our model not only allows the transmission rate of a connection to differ across the various connections, but also

a careful approach in formally defining throughput (see Section 2 for our precise definition).

A second difficulty in evaluating a protocol's throughput performance in the network model considered in this paper arises from the dynamic topology nature of the network. Indeed, for the model we will be considering, not only is the topology of the network not fixed, but we will *not* impose minimal connectivity requirements. In particular, since we are considering communication between two designated nodes through a network, there is no assumption made about how often (or even *if*) these two nodes are connected by a path. Indeed, we will model the dynamic nature of our network by introducing an *edge-scheduling* adversary who controls all the links in the network, and whose goal may be to disrupt communication as much as possible between the Sender and the Receiver. Since we make no connectivity assumptions, the adversary can simply choose to leave *all* the links inactive, rendering all communication impossible. We will handle this second difficulty by employing *competitive analysis*.

### 1.3 Our Results

In this paper, we lay the foundation for evaluating routing protocols in an "unrestricted" network model: on-line, distributed, asynchronous, dynamic, with bounded memory and no minimal connectivity assumptions. As demonstrated by the discussion above, network models that demonstrate some (but not all) of the weaker assumption choices have been considered by multiple authors, but to date no effort has been made to understand what (if any) performance guarantees are possible for extremely unreliable networks that are subject to all of the challenges modelled by the weak assumptions.

We focus on the task of end-to-end communication, and using competitive analysis we analyze the optimal throughput that can be achieved in this network model. We first demonstrate that the best possible competitive ratio that *any* protocol can hope to achieve is $1/n$. We then describe an explicit protocol that realizes this optimal competitive ratio. Informally, our results can be summarized:

**Theorem 1 (Informal)** *The best competitive-ratio that any protocol can achieve in a distributed asynchronous network with bounded-memory and dynamic topology (and no connectivity assumptions) is $1/n$. In particular, given any protocol $\mathcal{P}$, there exists an alternative protocol $\mathcal{P}'$, such that $\mathcal{P}'$ will out-perform $\mathcal{P}$ by a factor of at least $n$.*

**Theorem 2 (Informal)** *There exists a protocol that can achieve a competitive ratio of $1/n$ in a distributed asynchronous network with bounded-memory and dynamic topology (and no connectivity assumptions).*

Theorem 1 states that given any protocol $\mathcal{P}$, there exists an off-line protocol $\mathcal{P}'$ and a scheduling adversary such that $\mathcal{P}'$ will out-perform $\mathcal{P}$ by at least a factor of $n$. The proof of Theorem 1 (i.e. the lower bound) is highly nontrivial and relies on a delicate combinatorial argument together with a nonstandard potential function analysis (see further intuition in section 3).

Meanwhile, (the proof of) Theorem 2 exhibits a protocol "Slide+" that guarantees that even against an omniscient off-line protocol and against any adversary, Slide+ will never be out-performed by more than a factor of $n$. In other words, Slide+ is within a factor of $n$ of the highest possible throughput achievable by an omniscient algorithm that makes optimal routing decisions based on a complete view of the future conditions of the network and its topology. Moreover, by Theorem 1, this is the optimal guarantee for throughput that a protocol can hope to enjoy, i.e. no on-line strategy can achieve a better competitive ratio.

## 2 The Model

In this section, we describe formally the model in which we will be analyzing routing protocols. We begin by modeling the network as a graph $G$ with $n$ vertices (or *nodes*). Two of these nodes are

---

allows the same connection to have different transmission rates for each packet it transfers.

designated as the *Sender* and *Receiver*, and the Sender has a stream of packets $\{p_1, p_2, \dots\}$ that it wishes to transmit through the network to the Receiver.

Asynchronous communication networks vary from synchronous networks in that the transmission time across an edge in the network is not fixed (even along the same edge, from one message transmission to the next). Since there is no common global clock or mechanism to synchronize events, an asynchronous network is often said to be "message driven," in that the actions of the nodes in the network occurs exactly (and only) when they have just sent/received a message.

For this reason, asynchronous networks are commonly modelled by introducing an *edge-scheduling adversary* that controls the edges of the network as follows. A round is defined to be a single edge $E(u, v)$ in the network chosen by the adversary in which two sequential events occur: 1) Among the packets from $u$ to $v$ (and vice-versa) that the adversary is storing, it will choose one (in any manner it likes) and deliver it to $v$ (resp. to $u$);[8] 2) After seeing the delivered packet, $u$ (resp. $v$) sends requests of the form $(u, w, p) = $ (sending node, target node, packet) to the adversary, which will be stored by the adversary and may be delivered the *next* time $u$ (resp. $v$) is part of a round. Modelling asynchronicity in this manner captures both the intuition that a node has no idea how long a message that it "sends" to adjacent node $v$ will take to arrive, and it also considers "worst-case" asynchronicity in that a (potentially deliberately adversarial) adversary controls the scheduling of rounds/edges.

Aside from obeying the above specified rules, we place no restriction on the scheduling adversary. In other words, it may honor whatever edges it likes (this models the fact our network makes no connectivity assumptions), wait indefinitely long between honoring the same edge twice (modeling both the dynamic and asynchronous features of our network), and do anything else it likes (while respecting the guidelines) in attempt to hinder the performance of a routing protocol.

Note that our network model is on-line and distributed, in that we do not assume that the nodes have access to any information (including future knowledge of the adversary's schedule) aside from the packets they receive during a round they are a part of. Finally, we insist that nodes have bounded memory[9] $C$ (at most polynomial in $n$).

The goal of this paper is to analyze the performance of routing protocols in a network model that is: on-line, distributed, asynchronous, dynamic, with bounded memory and no connectivity assumptions. Our mechanism for evaluating protocols will be to measure their *throughput*, a notion we can now define formally in the context of rounds and the edge-scheduling adversary. In particular, let $f_{\mathcal{P}}^{\mathcal{A}} : \mathbb{N} \to \mathbb{N}$ be a function that measures, for a given protocol $\mathcal{P}$ and adversary $\mathcal{A}$, the number of packets that the Receiver has received as a function of the number of rounds that have passed.[10] The function $f_{\mathcal{P}}^{\mathcal{A}}$ formalizes our notion of throughput.

As mentioned in the Introduction, we utilize competitive analysis to gauge the performance (with respect to throughput) of a given protocol against all possible competing protocols. In particular, for any fixed adversary $\mathcal{A}$, we may consider the ideal "off-line" protocol $\mathcal{P}'$ which has perfect information.[11] That is, for any fixed round $x$, there exists an ideal off-line protocol $\mathcal{P}'(\mathcal{A}, x)$ such that $f_{\mathcal{P}'}^{\mathcal{A}}(x)$ is maximal.

**Definition 2.1.** We say that a protocol $\mathcal{P}$ has competitive ratio $1/\lambda_{\mathcal{P}}$ (respectively is $\lambda_{\mathcal{P}}$-competitive) if there exists a constant $k$ and function $g(n, C)$ such that for all possible adversaries $\mathcal{A}$ and for all $x \in \mathbb{N}$:

$$f_{\mathcal{P}'}^{\mathcal{A}}(x) \leq (k \, \lambda_{\mathcal{P}}) \cdot f_{\mathcal{P}}^{\mathcal{A}}(x) + g(n, C) \qquad (1)$$

Note that while $g$ may depend on the size of the network $n$ and the bounds placed on processor memory $C$, both $g$ and $k$ are *independent* of the round $x$ and the choice of adversary $\mathcal{A}$. We may now restate our two main results formally:

---

[8] For ease of discussion, we assume that all edges in the network have a fixed bandwidth/capacity, and that this quantity is the same for all edges in the network. We emphasize that this assumption does not restrict the validity of our claims in a more general model allowing varying bandwidths, but is only made for ease of exposition.

[9] For simplicity, we assume that all nodes have the same memory bound, although our argument can be readily extended to handle the more general case.

[10] In this paper, we will consider only deterministic protocols, so $f_{\mathcal{P}}^{\mathcal{A}}$ is well-defined.

[11] Here, perfect information means that the off-line protocol has knowledge of all future decisions of the adversary.

**Theorem 2.2.** *For any protocol $\mathcal{P}$ operating in a distributed, asynchronous, bounded memory network with dynamic topology (and no connectivity assumptions), the competitive ratio of $\mathcal{P}$ is:*

$$\lambda_\mathcal{P} \geq n$$

**Theorem 2.3.** *In a distributed, asynchronous, bounded memory network with dynamic topology (and no minimal connectivity assumptions), there exists a protocol $\mathcal{P}$ that is n-competitive ($\lambda_\mathcal{P} = n$).*

We prove Theorem 2.2 in the next section, and then go on to demonstrate a protocol in Section 4 that achieves competitive ratio $1/n$.

## 3 Optimal Competitive Ratio in Unrestricted Networks

Due to space constraints and the complexity of the argument, we will only be able to sketch the proof of Theorem 2.2 in this section. At a high level, the idea is to describe an adversary that schedules edges based on the given protocol's actions such that the packets of the protocol get "spread out" among the nodes of the network. Meanwhile, with knowledge of the adversary's schedule, an offline protocol can choose to only move packets along edges leading to the receiver. A short description is below; the full proof can be found in Appendix A.

The network model assumes that nodes have bounded memory, so let $C$ denote the maximal number of packets that any node can store at any time. We will show that for any deterministic protocol $\mathcal{P}$, there exists an adversary $\mathcal{A}$, a protocol $\mathcal{P}'$, and a sequence of strictly positive integers $\{m_1, m_2, \dots\}$ such that for any $\alpha > 0$, by round $x = \sum_{i=1}^{\alpha} m_i C$:

$$f_{\mathcal{P}'}^\mathcal{A}(x) = \alpha C \quad \text{and} \quad f_\mathcal{P}^\mathcal{A}(x) \leq \frac{\alpha C}{(n-2)} \approx \frac{\alpha C}{n}, \tag{2}$$

from which we conclude that the competitive ratio of $\mathcal{P}$ is at best $1/n$.

We begin by describing the adversary, i.e. a schedule (or order) of edges that will be honored. The schedule will proceed in **cycles**, with the $i^{th}$ cycle lasting $m_i C$ rounds. For the first $C$ rounds, the adversary finds the *internal* node $A_1$ that is currently storing the *most* packets (ties are broken arbitrarily), and honors edge $E(S, A_1)$ for $C$ rounds (here $S$ denotes the Sender). The protocol then proceeds inductively, starting with $j = 2$ and $\widehat{A}_1 = A_1$:

1. The adversary finds node $A_j$, where $A_j$ is the node in the network closest in height (but *smaller*) to $\widehat{A}_{j-1}$. If there is no such node, set $A_j$ to the Receiver $R$.
2. The adversary honors edge $E(\widehat{A}_{j-1}, A_j)$ for $C$ rounds
3. The adversary sets $\widehat{A}_j$ to be whichever node ($\widehat{A}_{j-1}$ or $A_j$) has *fewer* packets after the $C$ rounds of edge $E(\widehat{A}_{j-1}, A_j)$ has just passed.

The above three steps are continued until the end of the $C$ rounds for which $A_j = R$.

Notice a few features of the adversarial strategy: 1) The Sender's ability to insert packets is hindered by the fact the adversary is choosing to honor edge $E(S, N)$ for the node $N$ with the smallest capacity to store *more* packets; 2) By selecting in Step 2 the node storing fewer packets, the adversary is attempting to minimize the number of packets that make progress towards the Receiver; indeed 3) Among all nodes in the network, the node $N$ that is currently storing the fewest packets will be the one connected to the Receiver in the final $C$ rounds of the cycle. Also, it is clear that an off-line protocol $\mathcal{P}'$ with knowledge of all future rounds will be able to deliver $C$ packets every cycle. Since a cycle consists of $C * m$ rounds for some positive integer $m$, we can generate a sequence of positive integers $\{m_i\}$ coming from the $i^{th}$ cycle, yielding the first equality of (2), so it remains to prove the second bound in (2).

Fix any on-line protocol $\mathcal{P}$ we wish to analyze. If we could demonstrate that $\mathcal{P}$ delivers at most $C/(n-2)$ packets per cycle, then (2) would be immediate. Unfortunately, one can imagine e.g. the state of the network at the beginning of some cycle being such that *all* internal nodes are storing the maximum $C$ allowed packets. In this case, $\mathcal{P}$ will be able to deliver $C$ packets this round. Therefore, we instead need to argue that if $\mathcal{P}$ ever reaches a state where it is able to deliver more than $C/(n-2)$ packets in some cycle (e.g. all nodes are full), then it must be that $\mathcal{P}$ has delivered *fewer* than an average of $C/(n-2)$ packets per cycle in the past.

With this counter-example in mind, we define a potential function $\Psi^\alpha$, which intuitively measures the ability of $\mathcal{P}$ to deliver packets in the $\alpha^{th}$ cycle. We will show that whenever $\mathcal{P}$ delivers more than $C/(n-2)$ packets, the difference $\Psi^\alpha - \Psi^{\alpha+1}$ will be positive and "sufficiently large." Conversely, any time $\Psi^{\alpha+1} > \Psi^\alpha$, we will show that necessarily $\mathcal{P}$ delivered "significantly fewer" than $C/(n-2)$ packets in the $\alpha^{th}$ cycle.

Formally, for any $\alpha \in \mathbb{N}$, let $H_i^\alpha$ denote the number of packets that node $N_i^\alpha$ is storing at the outset of $\alpha$, and then define:

$$\Psi^\alpha = \sum_{i=1}^{n-2} \left(\frac{1}{2}\right)^{n-i-2} \cdot \max\left(0, H_i^\alpha - (n-i-2)\frac{C}{n-2}\right) \tag{3}$$

Let $Z^\alpha$ denote the number of packets the Receiver receives in the $\alpha^{th}$ cycle. Our main technical result for this section is then:

**Theorem 3.1.** *For all $\alpha \in \mathbb{N}$:*

$$Z^\alpha + (\Psi^{\alpha+1} - \Psi^\alpha) \leq \frac{7C}{n-2} \tag{4}$$

*Proof.* See the proof of Theorem A.11 in the Appendix.

With Theorem 3.1 in hand, we obtain the second inequality of (2) as an immediate corollary:

**Theorem 3.2.** *For any $\alpha \in \mathbb{N}$ and $x = (n-2)\alpha C$:*

$$f_\mathcal{P}^\mathcal{A}(x) \leq \frac{7\alpha C}{n-2} \tag{5}$$

*Proof.* Consider the string of inequalities:

$$f_\mathcal{P}^\mathcal{A}(x) = \sum_{\beta \leq \alpha} Z^\beta \leq \sum_{\beta \leq \alpha} \left(\frac{7C}{(n-2)} - (\Psi^{\beta+1} - \Psi^\beta)\right) = \frac{7\alpha C}{n-2} + \Psi^1 - \Psi^{\alpha+1} \leq \frac{7\alpha C}{n-2}, \tag{6}$$

where the last inequality follows from the fact that $\Psi^{\alpha+1} \geq 0$ and $\Psi^1 = 0$ (the latter is true since at the outset of the protocol, all nodes are not storing any packets).

## 4 Optimal On-line Local Control Protocol

In this section we present an on-line protocol that enjoys competitive ratio $1/n$. The protocol is a basic implementation of the "Slide" protocol (or *gravitational-flow*), which was first introduced by Afek and Gafni [2], and further developed in a series of work [14], [1], and [27]. We chose to analyze the performance of this protocol in our "unrestricted" network model because its inherent message-driven protocol is well-suited for the asynchronous network, and it has also been shown to out-perform more naive candidates for asynchronous routing protocols (e.g. broadcast) when stronger network assumptions are made [19].

In the following sub-section, we outline the basic Slide protocol, and then in Section 4.2 we sketch the proof that guarantees that the basic version of Slide is at least $n$-competitive in a "semi-asynchronous" network (defined below). In Appendix B, we present a modification of the Slide protocol that achieves the optimal $\frac{1}{n}$-competitive ratio in the fully asynchronous model of Section 2.

Recall that in the asynchronous model defined in Section 2, the adversary maintains a buffer of requests $(s, t, p)$ that it has received so far, and then the protocol proceeds in rounds where an edge $E(u, v)$ is honored, during which:

1. From its buffer, the adversary chooses one packet $(u, v, p_i)$ to deliver to $v$, and one packet $(v, u, p_j)$ to deliver to $u$
2. Based only on packets each node $u$ and $v$ has received thus far (and the instructions from the protocol), $u$ and $v$ send requests to the adversary

The basic Slide protocol dictates that a packet should be sent from $u$ to $v$ if $u$ *currently* has more packets stored than $v$. Therefore, a request $(u, v, p)$ that a node $u$ submits to the adversary will have additional information piggy-backed onto the packet $p$. Namely, $u$ will append its height (a number $H$ representing the number of packets $u$ currently holds) to the packet when submitting a request to the adversary, thus requests will have form $(u, v, (p, H))$.

Unfortunately, in the model above, when $u$ submits a request $(u, v, (p, H))$ to the adversary, the value for $H$ may become out-dated by the time the adversary honors edge $E(u, v)$ (for instance if the adversary honors edges $E(u, v')$ for $v' \neq v$ in between honors to $E(u, v)$). In the Appendix, we handle this problem by having nodes communicate "approximate" heights, and show that this approximation to the Slide protocol (which, together with several other technical modifications, we call Slide+) will not perform much differently than a protocol in which nodes have perfect information about their neighbors' heights when they are making a decision to send/receive a packet.[12] In the remainder of this section, we assume a semi-asynchronous model, defined as follows:

1'. The adversary does not maintain a buffer of requests of packets from nodes and must instead satisfy them immediately as specified in 3' below
2'. The adversary proceeds in the same manner as before, by selecting an edge $E(u, v)$ to honor according to the same guidelines as in Section 2
3'. During a round $E(u, v)$, the adversary first "awakens" $u$ and $v$ to alert them they are a part of the current round. Nodes $u$ and $v$ may now submit their request, consisting only of a packet plus control information, to the adversary who must directly deliver the packet $p$ to $v$ during this round (similarly the packet $p'$ that $v$ submitted is delivered to $u$).

We note that in terms of 3', it will not be important for the Slide protocol whether or not $u$ is aware of which edge the adversary is honoring, i.e. which node $v$ will be receiving $u$'s packet. Note that in the semi-asynchronous model, the problem of out-dated height information being transferred between two nodes is avoided. This modification is necessary to prove that the basic Slide protocol achieves competitive ratio $1/n$, but we emphasize that our result in Theorem 2.3 remains valid in the full asynchronous network model of Section 2. In the Appendix, we present the Slide+ protocol, and demonstrate how to extend the proof of competitive ratio $1/n$ for basic Slide in the semi-asynchronous model to an equivalent ratio for Slide+ in the fully asynchronous network model of Section 2.

### 4.1 Description of the Protocol

There are numerous instantiations of the Slide protocol that vary slightly between one another, but the basic principle is always the same. Due to space constraints, we will not provide a detailed description of the protocol, but refer the reader to [2] for the original protocol, and [14], [1], [27], and [19] for various modifications. Below, we present a basic implementation of the Slide protocol, and then go on to prove that the basic Slide protocol achieves competitive ratio $1/n$ in the restricted *semi-asynchronous* model of $1' - 3'$ described above.[13]

The network model assumes that nodes have bounded memory, so let $C$ denote the maximal number of packets that any node can store at any time.[14] Also, we will assume $C/n \in \mathbb{N}$ and in particular that $C/n \geq 2$ (the former assumption is not necessary but will make the exposition easier;

---
[12] Although the approximated protocol demands that nodes have buffer sizes at least $O(n^2)$, whereas normal Slide only requires buffers of size $O(n)$

[13] Somewhat surprisingly, even though the Slide protocol has been in existence for over a decade, no throughput competitive analysis for the asynchronous (or even semi-asynchronous) model has ever been performed.

[14] For simplicity, we assume that all nodes have the same memory bound, although our argument can be readily extended to handle the more general case.

the latter is necessary for the Slide protocol to work). Within the context of the semi-asynchronous network model ($1' - 3'$ above), we describe the request that a node $u$ will make to the adversary when it is "awakened," and also how this node $u$ will respond to the packet it receives from $v$:

1. If $u$ is the Sender, then $u$ finds the next packet $p_i \in \{p_1, p_2, \dots\}$ that has not yet been deleted (see 1a below), and forms: $p := (p_i, C + \frac{C}{n} - 1)$. Meanwhile, when $u$ receives (in the same round) the packet $(p_j, h)$:
    (a) If $h < C$, then $u$ deletes packet $p_i$ from his input stream $\{p_1, p_2, \dots\}$ (and ignores the received packet $p_j$)
    (b) If $h \geq C$, then $u$ keeps $p_i$ (and ignores the received packet $p_j$)

2. If $u$ is the Receiver, then $u$ forms the packet to send $p := (\bot, \frac{-C}{n})$. Meanwhile, when $u$ receives a packet of form $(p_j, h)$, if $p_j \neq \bot$, $u$ stores/outputs $p_j$ as a packet successfully received.

3. If $u$ is an internal node (not Sender or Receiver) and $u$ currently has height $H$, then $u$ finds the last packet $p_i$ that it has received, and sets: $p := (p_i, H)$ (if $H = 0$, then set $p_i = \bot$). Meanwhile, when $u$ receives (in the same round) a packet of form $(p_j, h)$:
    (a) If $H \geq h + C/n$, then $u$ will delete $p_i$ (and ignore the packet $p_j$)
    (b) If $H \leq h - C/n$, then $u$ will keep $p_i$, and also store $p_j$ (as the most recent packet received)
    (c) If $|H - h| < C/n$, then $u$ will keep $p_i$ and ignore packet $p_j$

Notice that rules 1-3 essentially state that internal nodes will always accept packets from the Sender (if they have room), always send packets to the Receiver (if they have any to send), and will transfer a packet to a neighboring internal node if and only if they are currently storing at least $C/n$ more packets than that neighbor.

### 4.2 Competitive Analysis of Slide in the Semi-Asynchronous Model

Due to space constraints, we provide here only a sketch of the proof that the above described Slide protocol enjoys competitive ratio $1/n$. The full proof can be found in Appendix B.

Recall that we wish to show that there exists a constant $k$ and function $g(n, C)$ such that for any round $x$ and against any adversary $\mathcal{A}$ (see (1)):

$$f^{\mathcal{A}}_{\mathcal{P}'}(x) \leq (kn) \cdot f^{\mathcal{A}}_{\mathcal{P}}(x) + g(n, C) \qquad (7)$$

Above (and through the remainder of this section), $\mathcal{P}$ will denote the Slide protocol, and for fixed choice of adversary $\mathcal{A}$ and round $x$, $\mathcal{P}'(\mathcal{A}, x)$ will denote the ideal off-line protocol (since we will be fixing $x$ and $\mathcal{A}$, we will usually write simply $\mathcal{P}'$). We will show that (7) will be true for all rounds $x$ and all adversaries $\mathcal{A}$ for $k = 4$ and $g(n, C) = 4n^2 C$. We proceed by fixing an arbitrary adversary $\mathcal{A}$ and round $x \in \mathbb{N}$, and showing that for these (arbitrary) choices, (7) will be satisfied.

We begin with some notation and terminology. Any time the Sender reaches Step 1a, we will say packet $p_i$ was inserted. Similarly, anytime the Receiver stores/outputs packet $p_j$, we will say the packet has been received. Notice that anytime an internal node $u$ reaches Step 3a, the node $v$ at the other end has necessarily reached Step 3b. In this case, we will say packet $p_i$ was transferred from $u$ to $v$. Let $Y^{\mathcal{P}}$ denote the number of packets inserted by $\mathcal{P}$ as of round $x$ ($Y^{\mathcal{P}'}$ is defined analogously). The starting point of the proof is the following trivial observation:

**Observation 1.** *As of round $x$, there have been at most $2nY^{\mathcal{P}}$ packet transfers in $\mathcal{P}$.*

The observation is a product of the fact that packets are transferred in a FILO manner, so rules 3a-3c state that a packet must drop in *height* by at least $C/n - 1$ when it is transferred. Since a node's memory is $C \geq 2n$, a single packet can be transferred at most $2n$ times.

Let $Z^{\mathcal{P}'}$ denote the packets that have been received by the Receiver for protocol $\mathcal{P}'$ as of round $x$ (define $Z^{\mathcal{P}}$ analogously). Notice that $f^{\mathcal{A}}_{\mathcal{P}'}(x)$, the left-hand-side of (7), is equal to $|Z^{\mathcal{P}'}|$ (we will occasionally write $Z^{\mathcal{P}'}$ when we really mean $|Z^{\mathcal{P}'}|$; the meaning will be clear from context). We split $Z^{\mathcal{P}'}$ into two disjoint subsets $Z^{\mathcal{P}'} = Z_1^{\mathcal{P}'} \cup Z_2^{\mathcal{P}'}$, which we now describe.

We can view the adversary $\mathcal{A}$ as simply a schedule (or order) of edges that the adversary will honor. We will imagine a virtual world, in which the two protocols (Slide and the ideal off-line protocol) are run simultaneously in the same network. Define $Z_1^{\mathcal{P}'}$ to be the subset of $Z^{\mathcal{P}'}$ consisting of packets $p'$ for which there exists at least one round $E(u,v)$ such that both $p'$ *and* some packet $p \in Y^{\mathcal{P}}$ were both transferred this round.[15] Set $Z_2^{\mathcal{P}'} = Z^{\mathcal{P}'} \setminus Z_1^{\mathcal{P}'}$.

**Lemma 4.1.** $\quad |Z_1^{\mathcal{P}'}| \leq 2n|Z^{\mathcal{P}}| + 2n^2 C$

*Proof.* This follows from Observation 1 together with the fact $|Y^{\mathcal{P}} - Z^{\mathcal{P}}| \leq nC$.

It remains to bound $|Z_2^{\mathcal{P}'}| \leq 2n|Z^{\mathcal{P}}| + 2n^2 C$. To this end, we first observe that when any packet $p' \in Z_2^{\mathcal{P}'}$ was first inserted, it was necessarily inserted into some node $u$ such that *with respect to* $\mathcal{P}$, $u$ had height $C$ (otherwise $\mathcal{P}$ would have also inserted a packet this round, and then $p' \in Z_1^{\mathcal{P}'}$). Similarly, when $p' \in Z_2^{\mathcal{P}'}$ is received by the Receiver from some node $v$, then *with respect to $\mathcal{P}$*, $v$ had height zero. The idea will then be to assign a potential function $\varphi_{p'}$ to every packet $p' \in Z_2^{\mathcal{P}'}$ that represents the current height *with respect to* $\mathcal{P}$ of the node in which $p'$ is currently stored. Thus, when a packet $p' \in Z_2^{\mathcal{P}'}$ is first inserted, $\varphi_{p'} = C$, and when $p' \in Z_2^{\mathcal{P}'}$ is received, $\varphi_{p'} = 0$.

Next, we define a second potential function $\Phi_{\mathtt{t}}$, which will obey:

1. $\Phi_0 = 0$ at the outset of the protocol, and every time there is a packet transfer in $\mathcal{P}$, $\Delta\Phi = C$
2. For any packet $p' \in Z_2^{\mathcal{P}'}$, anytime $\varphi_{p'}$ changes (aside from when $\varphi_{p'}$ is initialized to $C$ upon insertion of $p'$), $\Phi$ changes by an equivalent amount

With these definitions, we have:

**Lemma 4.2.** *For any round $\mathtt{t} \leq x$, let $\mathcal{Z}_{\mathtt{t}} \subseteq Z_2^{\mathcal{P}'}$ denote the set of packets in $Z_2^{\mathcal{P}'}$ that have been inserted by round $\mathtt{t}$. Then:*

$$\Phi_{\mathtt{t}} = C * (\textit{No. of transfers in } \mathcal{P} \textit{ as of } \mathtt{t}) \; - \sum_{p' \in \mathcal{Z}_{\mathtt{t}}} (C - \varphi_{p'})$$

$$\leq C * \left[ (\textit{No. of transfers in } \mathcal{P} \textit{ as of } \mathtt{t}) - |Z_2^{\mathcal{P}'}| \right]$$

$$\leq C * (2nY^{\mathcal{P}} - |Z_2^{\mathcal{P}'}|) \tag{8}$$

Consequently, if we can show that at all times $\Phi \geq 0$, (8) implies:

$$|Z_2^{\mathcal{P}'}| \leq 2nY^{\mathcal{P}} \leq 2n|Z^{\mathcal{P}}| \; + \; 2n^2 C \tag{9}$$

Thus, provided $\Phi \geq 0$, we can conclude $|Z^{\mathcal{P}'}| = |Z_1^{\mathcal{P}'}| + |Z_2^{\mathcal{P}'}| \leq 4n|Z^{\mathcal{P}}| \; + \; 4n^2 C$, as required. The main technical challenge is to argue why $\Phi \geq 0$, which is the content of Lemma B.17, and will require a bit of work, all of which can be found in Appendix B.

## 5 Conclusion

In this paper, we investigated the feasibility of routing in a distributed, asynchronous, bounded memory, network with dynamic topology and no minimal assumptions on connectivity. In particular, we used competitive analysis to evaluate optimal throughput performance of end-to-end communication routing protocols in this general network. Within this setting, our first result was to prove a bound of $1/n$ as the best-possible throughput of any deterministic protocol. That is, for any protocol, there exists a competing protocol together with a schedule of active edges, such that the competing protocol will out-perform the given protocol (in terms of throughput) by at least a factor of $n$.

---

[15] Note that we make no condition that the two packets traveled in the same direction.

We then went on to demonstrate that in the *semi-asynchronous* network model, the Slide protocol achieves the optimal competitive ratio of $1/n$. In Appendix C, we present the "Slide+" protocol, which is a modification of the standard Slide protocol that allows us to achieve the same optimal competitive ratio of $1/n$ in the fully asynchronous model. By Theorem 2.2, this is the optimal guarantee for throughput that a protocol can hope to enjoy in the general network model presented in Section 2.

# Appendix

## A  Formal Proof of Throughput Bound

In this section, we go through the rigorous details of the proof of Theorem 2.2, which was sketched in Section 3. We will use the same notation introduced there for the remainder of this section. In particular, recall that there is some fixed protocol $\mathcal{P}$ that we wish to analyze, and we are considering a scheduling adversary $\mathcal{A}$ that proceeds in cycles.

We begin with a reduction of the given protocol $\mathcal{P}$ to a *virtual* protocol $\mathcal{P}'$, which will be operating with respect to a *different* scheduling adversary $\mathcal{A}'$ than $\mathcal{P}$. The schedule of edges honored by $\mathcal{A}'$ will be (in general) different than those honored by $\mathcal{A}$, but $\mathcal{A}'$ will also proceed in *cycles*. For any cycle $\alpha$ in $\mathcal{P}'$'s world, define $\Psi'^\alpha$ and $Z'^\alpha$ analogous to $\Psi^\alpha$ and $Z^\alpha$ that were defined for $\mathcal{P}$ in Section 3. We emphasize that the two worlds of $\mathcal{P}$ and $\mathcal{P}'$ are different, and we are *not* attempting to apply competitive analysis to these two protocols. Rather, the property that $\mathcal{P}'$ will satisfy is:

$$\forall \alpha \in \mathbb{N}: \quad \Psi^\alpha = \Psi'^\alpha \quad \text{and} \quad Z^\alpha = Z'^\alpha \tag{10}$$

Then given that (10) holds for all cycles $\alpha$, if we can show for all $\alpha$ (subject to $\mathcal{A}'$'s schedule):

$$Z'^\alpha + (\Psi'^{\alpha+1} - \Psi'^\alpha) \leq \frac{7C}{n-2}, \tag{11}$$

then the equivalent statement will be true for $\mathcal{P}$, which is Theorem 3.1 in Section 3, and thus the proof will be complete.

We now explain the alternate scheduling adversary $\mathcal{A}'$, which will be defined in terms of any arbitrary protocol attempting to route in a network controlled by $\mathcal{A}'$. As mentioned above, the schedule of $\mathcal{A}'$ will proceed in cycles, each of which will last $(n-1)C$ rounds. At the beginning of any cycle $\alpha$, $\mathcal{A}'$ labels the internal nodes by $\{N_1^\alpha, N_2^\alpha, \ldots, N_{n-2}^\alpha\}$, so that for all $1 \leq i \leq n-3$, node $N_i^\alpha$ is storing *more* packets than $N_{i+1}^\alpha$ at the outset of cycle $\alpha$ (note that the labels/indices of the internal nodes will change every cycle). For the first $C$ rounds of the cycle, the adversary will honor edge $E(S, N_1)$ (here $S$ denotes the Sender). We describe the remaining rounds in this cycle inductively (starting below for $i = 1$, and $\widetilde{N}_1^\alpha = N_1^\alpha$):

1. The adversary honors edge $E(\widetilde{N}_i^\alpha, N_{i+1}^\alpha)$ for $C$ rounds

2. After the first $(i+1)C$ rounds of cycle $\alpha$ have passed (i.e. edge $E(\widetilde{N}_i^\alpha, N_{i+1}^\alpha)$ has just been honored $C$ times), let $\widetilde{N}_{i+1}^\alpha \in \{\widetilde{N}_i^\alpha, N_{i+1}^\alpha\}$ denote the node storing *fewer* packets than the other.

Steps 1-2 are repeated through $i = n-1$, so that $E(\widetilde{N}_{n-1}^\alpha, N_{n-2}^\alpha)$ has just completed, and $\widetilde{N}_{n-2}^\alpha$ has been defined. Then for the last $C$ rounds of cycle $\alpha$, the adversary honors edge $E(\widetilde{N}_{n-2}^\alpha, R)$.

**Lemma A.1.** *Given protocol $\mathcal{P}$ routing in a network controlled by $\mathcal{A}$ (whose schedule was described in Section 3), there exists a protocol $\mathcal{P}'$ competing against $\mathcal{A}'$, such that with respect to each protocol's own cycle, (10) is valid.*

*Proof.* Since we are considering only deterministic protocols, we can define what $\mathcal{P}'$ will do in any round based on what $\mathcal{P}$ is doing. We will actually demonstrate something slightly stronger than (10), that is:

> **Induction Hypothesis.** Up to permutation of the internal nodes, the heights of each of the internal nodes in both worlds is the same at the start/end of any cycle, as is the number of packets delivered in any cycle.

We proceed by induction on the cycle. In particular, fix some cycle $\alpha$, and assume that the induction hypothesis is true for all cycles $\beta < \alpha$. In the first $C$ rounds of $\alpha$ in $\mathcal{P}$'s world, $\mathcal{A}$ opens edge $E(S, A_1)$, where $A_1$ is the internal node currently storing the most packets. Similarly, in the first $C$ rounds, $\mathcal{A}'$ opens edge $E(S, A'_1)$, where $A'_1$ is the internal node currently storing the most packets in $\mathcal{P}'$'s world. By the induction hypothesis, although the labels of node $A_1$ verses $A'_1$ may be different, the node that label represents will have the same height in the two worlds, and we define $\mathcal{P}'$ to do the same thing that $\mathcal{P}$ does in these first $C$ rounds.

Let $A_2$ denote the node for which the adversary $\mathcal{A}$ will honor edge $E(A_1, A_2)$ for the next $C$ rounds, and similarly for $A'_2$ with respect to $\mathcal{A}'$. Note that by the induction hypothesis together with the definition of $\mathcal{P}'$ (so far) for the first $C$ rounds of cycle $\alpha$, we have that the height of $A_1$ equals the height of $A'_1$, and similarly the heights of $A_2$ and $A'_2$ match. Now define $\mathcal{P}'$ to do in the $C$ rounds $E(A'_1, A'_2)$ whatever $\mathcal{P}$ does in the $C$ rounds $E(A_1, A_2)$.[16] Thus, after $2C$ rounds have passed, the two networks are still identical (up to permutation of the nodes).

Let $\widetilde{A}_2$ denote the node among $\{A_1, A_2\}$ that is storing fewer packets after the $C$ rounds of $E(A_1, A_2)$. Now in $\mathcal{P}$'s world, the adversary will search for the node $A_3$ with height closest to (but *smaller* than) $\widetilde{A}_2$, and the adversary $\mathcal{A}$ will next honor edge $E(\widetilde{A}_2, A_3)$ for $C$ rounds. Notice that, if e.g. $\mathcal{P}$ had $A_2$ transfer all its packets to $A_1$ during the $C$ rounds of $E(A_1, A_2)$, it is possible that $A_3$ is *not* the node that had the third highest height at the start of cycle $\alpha$ (indeed, its even possible that $A_3 = R$).

By the induction hypothesis, there is some node $A'_i$ ($i \geq 3$) in $\mathcal{P}'$'s world such that at the start of $\alpha$, the height of $A_3$ equals the height of $A'_i$ (if $A_3 = R$, then $i = n - 1$, i.e. set $A'_i = R$). Notice that in contrast to $\mathcal{P}$'s world, the schedule of $\mathcal{A}'$ will necessarily go through *every* internal node at least once. Indeed, for any $2 \leq m \leq n - 2$, the node in $\mathcal{P}'$'s world that started cycle $\alpha$ as the $m^{th}$ fullest node will necessarily be a part of rounds $mC$ through $(m+1)C - 1$. Therefore, for each $3 \leq m \leq i$, dictate that during rounds $mC$ through $(m+1)C - 1$, protocol $\mathcal{P}'$ will have the two nodes swap final states. In particular, for any $3 \leq m \leq i$, if $H'_m$ denotes the height of $A'_m$ at the start of cycle $\alpha$, then we dictate that $\mathcal{P}'$ has transfers enough packets from $A'_m$ to $A'_{m-1}$ during the $C$ rounds of $E(A'_{m-1}, A'_m)$ such that the height of $A'_{m-1}$ at the end of the $C$ rounds is equal to $H'_m$. In this manner, it is clear that by the time the virtual world of $\mathcal{P}'$ reaches the end of $iC$ cycles (recall that $i$ is defined so that the height of $A_3$ equals the height of $A'_i$), the state of the networks in the two worlds will be identical (up to permutation of the nodes). Furthermore, during the next $C$ rounds of each cycle, the adversaries $\mathcal{A}$ and $\mathcal{A}'$ will honor an edge between two nodes ($E(A_2, A_3)$ verses $E(A'_{i-1}, A'_i)$) such that at the moment the $C$ rounds start, the height of $A_2$ equals $A'_{i-1}$, and the height of $A_3$ equals $A'_i$. Therefore, this process may be repeated iteratively through the end of the cycle in each respective world, and it is clear that the induction hypothesis will remain valid by the end of cycle $\alpha$. ∎

For the remainder of the section, we will seek to prove (11) for the protocol $\mathcal{P}'$. To simplify notation, it will be convenient to define $m = n - 2$. At the outset of every cycle $\alpha$, we label the

---

[16] In order to preserve Fact 1 below, we demand that after the $C$ rounds of $E(A'_1, A'_2)$, $A'_2$ is storing *fewer* packets than $A'_1$. Therefore, if this is *not* the case for $E(A_1, A_2)$, then define $\mathcal{P}'$ to end in a symmetric state as $\mathcal{P}$, i.e. so that the pair of nodes $(A_1, A_2)$ have the same height as the pair of nodes $(A'_1, A'_2)$, but in the latter pair, necessarily $A'_1$ is storing at least as many packets as $A'_2$ after the $C$ rounds of $E(A'_1, A'_2)$.

internal (i.e. excluding the Sender and Receiver) nodes $\{N_1^\alpha, N_2^\alpha, \ldots, N_m^\alpha\}$, such that if $i < j$, then node $N_i^\alpha$ is storing *more* (or an equal number of) packets at the start of cycle $\alpha$ than $N_j^\alpha$. For all $\alpha$, let $N_0^\alpha = S$ and $N_{n-1}^\alpha = R$. For any $1 \leq i \leq n-2$, let $H_i^\alpha$ denote the height the node had *at the outset of* $\alpha$. We emphasize that while the heights of nodes may change through the course of cycle $\alpha$, the labeling $\{N_i^\alpha\}$ and the quantities $\{H_i^\alpha\}$ will remain fixed throughout the cycle. Indeed, the following fact implies that the labeling of nodes is independent of $\alpha$ (and in fact is fixed for all time):

**Fact 1.** For all $\alpha \in \mathbb{N}$ and all $1 \leq i \leq m$: $N_i^\alpha = N_i^{\alpha+1}$

**Fact 2.** For any cycle $\alpha$, node $N_i$ is a part of $2C$ rounds of the cycle: first for $C$ rounds with $E(N_{i-1}, N_i)$, and then for $C$ rounds with $E(N_i, N_{i+1})$

These facts, along with the following observations, all follow from the definition/construction of $\mathcal{P}'$ in the proof of Lemma A.1 above. To fix notation, for each $0 \leq i \leq m$ let $A_i^\alpha$ denote the number of packets sent from $A_i$ to $A_{i+1}$ during the $C$ rounds $E(N_i, N_{i+1})$ of cycle $\alpha$. Note that $A_i^\alpha$ may be negative if the net packet flow during $E(N_i, N_{i+1})$ was towards $N_i$.

**Lemma A.2.** *For any cycle $\alpha$ and for any $1 \leq i \leq m$:*

$$1)\ A_i^\alpha \leq \frac{A_{i-1}^\alpha + H_i^\alpha - H_{i+1}^\alpha}{2} \tag{12}$$

$$2)\ A_i^\alpha \leq H_i^{\alpha+1} - H_{i+1}^\alpha \tag{13}$$

*Proof.* Statement 1 follows from the two facts above as follows. Note that after the $C$ rounds $E(N_{i-1}, N_i)$ but before the next $C$ rounds, node $N_i$ will have height $A_{i-1}^\alpha + H_i^\alpha$. Now by definition of protocol $\mathcal{P}'$, at the end of the $C$ rounds of $E(N_i, N_{i+1})$, $N_i^\alpha$ will have a greater (or equal) number of packets than $N_{i+1}^\alpha$. In particular, since there are $A_{i-1}^\alpha + H_i^\alpha + H_{i+1}^\alpha$ total packets between the two nodes at the start of the $C$ rounds $E(N_i^\alpha, N_{i+1}^\alpha)$, it must be that at the end of these $C$ rounds, $N_i^\alpha$ is storing at least half of these. Since the number of packets stored by $N_i^\alpha$ after the $C$ rounds of $E(N_i^\alpha, N_{i+1}^\alpha)$ is given by $A_{i-1}^\alpha + H_i^\alpha - A_i^\alpha$, Statement 1 follows.

Also, again since protocol $\mathcal{P}'$ specifies that $N_i^\alpha$ must have more (or an equal number of) packets as $N_{i+1}^\alpha$ immediately after the $C$ rounds of $E(N_i^\alpha, N_{i+1}^\alpha)$, and by Fact 2 the height of $N_i^\alpha$ will not change through the remainder of cycle $\alpha$, Statement 2 follows. ∎

We are interested in the potential function:

$$\Psi'^\alpha = \sum_{i=1}^{m} \left(\frac{1}{2}\right)^{m-i} \cdot \max\left(0, H_i^\alpha - (m-i)\frac{C}{m}\right) \tag{14}$$

For each $1 \leq i \leq m$, define:

$$\delta_i^\alpha = \begin{cases} 1 & \text{if the } 2^{nd} \text{ term of the max statement in (14) dominates} \\ 0 & \text{otherwise} \end{cases} \tag{15}$$

Also, for any pair of indices $1 \leq j < k \leq m$, define:

$$(\Psi'^{\alpha+1} - \Psi'^\alpha)_{i,j} = \sum_{k=i}^{j} \left(\frac{1}{2}\right)^{m-k} \cdot \left[\max\left(0, H_k^{\alpha+1} - (m-k)\frac{C}{m}\right) - \max\left(0, H_k^\alpha - (m-k)\frac{C}{m}\right)\right] \tag{16}$$

**Claim A.3.** *For any index $1 \leq i \leq m$ and any cycle $\alpha$:*

$$H_i^{\alpha+1} = H_i^\alpha + A_{i-1}^\alpha - A_i^\alpha \tag{17}$$

*Proof.* Notice $N_i^{\alpha+1} = N^\alpha$ (Fact 1) and $N_i$ is a part of exactly $2C$ rounds for the $\alpha^{th}$ cycle (Fact 2). In the first $C$ rounds, $H_i$ changes by $A_{i-1}^\alpha$, and in the second $C$ rounds it changes by $-A_i^\alpha$. Since $N_i$ began the cycle with height $H_i^\alpha$, we have that its height at the start of the $(\alpha+1)^{th}$ cycle will be $H_i^\alpha + A_{i-1}^\alpha - A_i^\alpha$. ∎

It will be convenient to introduce the following notation:

**Definition A.4.** For any $1 \leq i \leq m$ and any cycle $\alpha$, define:

$$\omega_i^\alpha := \min\left(0, H_i^\alpha - (m-i)\frac{C}{m}\right) \tag{18}$$

**Claim A.5.** *For any index $1 \leq i \leq m$ and any cycle $\alpha$:*

$$1)\ \text{If } \delta^{\alpha+1} = 1,\ \text{then:}\quad (\Psi'^{\alpha+1} - \Psi'^\alpha)_{i,i} = \frac{1}{2^{m-i}}(A_{i-1}^\alpha - A_i^\alpha + \omega_i^\alpha)$$

$$2)\ \text{If } \delta^{\alpha+1} = 0,\ \text{then:}\quad (\Psi'^{\alpha+1} - \Psi'^\alpha)_{i,i} = \frac{1}{2^{m-i}}\omega_i^\alpha \tag{19}$$

*Proof.* If $\delta^{\alpha+1} = 1$, then consider the equalities:

$$(\Psi'^{\alpha+1} - \Psi'^\alpha)_{i,i} = \frac{1}{2^{m-i}}\left[\max\left(0, H_i^{\alpha+1} - (m-i)\frac{C}{m}\right) - \max\left(0, H_i^\alpha - (m-i)\frac{C}{m}\right)\right]$$

$$= \frac{1}{2^{m-i}}\left[(A_{i-1}^\alpha - A_i^\alpha + H_i^\alpha) - (m-i)\frac{C}{m} - \max\left(0, H_i^\alpha - (m-i)\frac{C}{m}\right)\right]$$

$$= \frac{1}{2^{m-i}}(A_{i-1}^\alpha - A_i^\alpha) + \begin{cases} 0 & \text{if } H_i^\alpha \geq (m-i)\frac{C}{m} \\ \frac{1}{2^{m-i}}\left(H_i^\alpha - \frac{(m-i)C}{m}\right) & \text{if } H_i^\alpha < (m-i)\frac{C}{m} \end{cases}$$

$$= \frac{1}{2^{m-i}}(A_{i-1}^\alpha - A_i^\alpha + \omega_i^\alpha)$$

where the second equality is from Claim A.3 together with the assumption that $\delta^{\alpha+1} = 1$. Otherwise, if $\delta^{\alpha+1} = 0$, then Statement 2 is immediate. ∎

**Lemma A.6.** *For any pair of indices $1 \leq i < j < m$ for which $\delta_k^{\alpha+1} = 1$ for every $i \leq k \leq j$:*[17]

$$(\Psi'^{\alpha+1} - \Psi'^\alpha)_{i,j} - \sum_{k=i}^{j} \frac{\omega_k}{2^{m-k}} + \frac{A_j}{2^{m-j-1}} \leq \frac{A_{i-1}}{2^{m-i}} + \frac{(j-i+1)}{2^{m-i+1}}(A_{i-1}+H_i) - \frac{H_{j+1}}{2^{m-j+1}} + \sum_{k=i+1}^{j-1} \frac{(j-k)}{2^{m-k+2}} H_k \tag{20}$$

*Proof.* This follows via an inductive argument on $j - i$ together with Lemma A.2 and Claim A.5:
<u>Base Case:</u> $j = i+1$: First consider the right-hand-side of (20) with $j = i+1$:

$$\text{RHS (20)} = \frac{A_{i-1}}{2^{m-i}} + \frac{2}{2^{m-i+1}}(A_{i-1}+H_i) - \frac{H_{i+2}}{2^{m-i}}$$

$$= \frac{A_{i-1}}{2^{m-i}} + \frac{1}{2^{m-i}}(A_{i-1}+H_i) - \frac{H_{i+2}}{2^{m-i}}$$

$$= \frac{A_{i-1}}{2^{m-i-1}} + \frac{1}{2^{m-i}}H_i - \frac{H_{i+2}}{2^{m-i}} \tag{21}$$

---
[17]On the right-hand side of (20), all superscripts are $\alpha$, which we have suppressed for notational convenience.

Meanwhile, for $j = i + 1$, the left-hand-side of (20) is:

$$\text{LHS (20)} = (\Psi'^{\alpha+1} - \Psi'^{\alpha})_{i,i+1} - \sum_{k=i}^{i+1} \frac{\omega_i}{2^{m-i}} + \frac{A_{i+1}}{2^{m-i-2}}$$

$$= (\Psi'^{\alpha+1} - \Psi'^{\alpha})_{i,i} + (\Psi'^{\alpha+1} - \Psi'^{\alpha})_{i+1,i+1} - \sum_{k=i}^{i+1} \frac{\omega_i}{2^{m-i}} + \frac{A_{i+1}}{2^{m-i-2}}$$

$$= \frac{1}{2^{m-i}}(A_{i-1} - A_i + \omega_i) + \frac{1}{2^{m-i-1}}(A_i - A_{i+1} + \omega_{i+1}) - \sum_{k=i}^{i+1} \frac{\omega_i}{2^{m-i}} + \frac{A_{i+1}}{2^{m-i-2}}$$

$$= \frac{1}{2^{m-i-1}} A_{i+1} + \frac{1}{2^{m-i}}(A_i + A_{i-1})$$

$$\leq \frac{1}{2^{m-i-1}}(A_i + H_{i+1} - H_{i+2}) + \frac{1}{2^{m-i}}(A_i + A_{i-1})$$

$$\leq \frac{1}{2^{m-i-1}}(H_i + A_{i-1} - A_i - H_{i+2}) + \frac{1}{2^{m-i}}(A_i + A_{i-1})$$

$$= \frac{A_{i-1}}{2^{m-i-1}} + \frac{1}{2^{m-i}} H_i - \frac{H_{i+2}}{2^{m-i}} \tag{22}$$

where the third equality is due to Claim A.5, the first inequality is Statement 1 of Lemma A.2, and the second inequality is Claim A.3. Notice (21) matches (22), as required.

INDUCTION STEP: Consider the string of inequalities:

$$(\Psi'^{\alpha+1} - \Psi'^{\alpha})_{i,j} - \sum_{k=i}^{j} \frac{\omega_i}{2^{m-i}} + \frac{A_j}{2^{m-j-1}} = (\Psi'^{\alpha+1} - \Psi'^{\alpha})_{i,i} + (\Psi'^{\alpha+1} - \Psi'^{\alpha})_{i+1,j} - \sum_{k=i}^{j} \frac{\omega_i}{2^{m-i}} + \frac{A_j}{2^{m-j-1}}$$

$$\leq \frac{1}{2^{m-i}}(A_{i-1} - A_i) + \frac{A_i}{2^{m-i-1}} + \frac{(j-i)}{2^{m-i}}(A_i + H_{i+1})$$

$$- \frac{H_{j+1}}{2^{m-j+1}} + \sum_{k=i+2}^{j-1} \frac{(j-k)}{2^{m-k+2}} H_k$$

$$\leq \frac{A_{i-1}}{2^{m-i}} + \frac{(j-i+1)}{2^{m-i+1}}(A_{i-1} + H_i) - \frac{H_{j+1}}{2^{m-j+1}} + \sum_{k=i+1}^{j-1} \frac{(j-k)}{2^{m-k+2}} H_k$$

where the first inequality is by the induction hypothesis together with Claim A.5 and the last inequality is by Statement 2 of Lemma A.2. ∎

**Lemma A.7.** *For any pair of indices $1 \leq i < i+1 < j \leq m$ for which $\delta_j^{\alpha+1} = 1$ but $\delta_k^{\alpha+1} = 0$ for every $i < k < j$:*[18]

$$(\Psi'^{\alpha+1} - \Psi'^{\alpha})_{i+1,j-1} - \sum_{k=i+1}^{j-1} \frac{\omega_k}{2^{m-k}} + \frac{A_{j-1}}{2^{m-j}} \quad \leq \quad \frac{A_i}{2^{m-i-1}} + \frac{H_{i+1}}{2^{m-i}} - \frac{H_j}{2^{m-j+1}} + \sum_{k=i+1}^{j-1} \frac{H_k}{2^{m-k+1}} \tag{23}$$

*Proof.* This follows via an inductive argument on $j - i$ together with Lemma A.2:

BASE CASE: $j - i = 2$: Looking at the right-hand-side of (23) for $j = i + 2$:

$$\text{RHS (23)} = \frac{A_i}{2^{m-i-1}} + \frac{H_{i+1}}{2^{m-i}} - \frac{H_{i+2}}{2^{m-i-1}} + \frac{H_{i+1}}{2^{m-i}}$$

$$= \frac{A_i + H_{i+1} - H_{i+2}}{2^{m-i-1}} \tag{24}$$

---

[18] On the right-hand side of (20), all superscripts are $\alpha$, which we have suppressed for notational convenience.

Meanwhile, looking at the left-hand-side of (23) for $j = i + 2$:

$$\text{LHS (23)} = (\Psi'^{\alpha+1} - \Psi'^{\alpha})_{i+1,i+1} - \sum_{k=i+1}^{i+1} \frac{\omega_i}{2^{m-i}} + \frac{A_{i+1}}{2^{m-i-2}}$$

$$= \frac{A_{i+1}}{2^{m-i-2}}$$

$$\leq \frac{A_i + H_{i+1} - H_{i+2}}{2^{m-i-1}}, \qquad (25)$$

where the second equality is from Claim A.5 (since $\delta_{i+1}^{\alpha+1} = 0$) and the inequality is Statement 1 of Lemma A.2.

INDUCTION STEP: Consider the string of inequalities:

$$(\Psi'^{\alpha+1} - \Psi'^{\alpha})_{i+1,j-1} - \sum_{k=i}^{j} \frac{\omega_i}{2^{m-i}} + \frac{A_{j-1}}{2^{m-j}} = (\Psi'^{\alpha+1} - \Psi'^{\alpha})_{i+1,i+1} + (\Psi'^{\alpha+1} - \Psi'^{\alpha})_{i+2,j-1} - \sum_{k=i}^{j} \frac{\omega_i}{2^{m-i}} + \frac{A_{j-1}}{2^{m-j}}$$

$$\leq \frac{A_{i+1}}{2^{m-i-2}} + \frac{H_{i+2}}{2^{m-i-1}} - \frac{H_j}{2^{m-j+1}} + \sum_{k=i+2}^{j-1} \frac{H_k}{2^{m-k+1}}$$

$$\leq \frac{A_i}{2^{m-i-1}} + \frac{H_{i+1}}{2^{m-i}} - \frac{H_j}{2^{m-j+1}} + \sum_{k=i+1}^{j-1} \frac{H_k}{2^{m-k+1}} \qquad (26)$$

where the first inequality is by the induction hypothesis together with Claim A.5 and the last inequality is by Statement 1 of Lemma A.2. ∎

**Lemma A.8.** *For any cycle $\alpha$ and any index $1 \leq i < m-1$, if $\delta_i^{\alpha+1} = 1$, $\delta_{i+1}^{\alpha+1} = 0$, and $\delta_{i+2}^{\alpha+1} = 1$, then:*

$$(\Psi'^{\alpha+1} - \Psi'^{\alpha})_{i+1,i+1} - \sum_{k=i+1}^{i+1} \frac{\omega_k}{2^{m-k}} + \frac{A_i}{2^{m-i-1}} \leq \frac{A_i}{2^{m-i-1}} + \frac{1}{2^{m-i-1}} \frac{C}{m} \qquad (27)$$

*Proof.* Consider:

$$(\Psi'^{\alpha+1} - \Psi'^{\alpha})_{i+1,i+1} - \sum_{k=i+1}^{i+1} \frac{\omega_k}{2^{m-k}} + \frac{A_i}{2^{m-i-1}} = \frac{A_{i+1}}{2^{m-i-2}}$$

$$\leq \frac{A_i + H_i - H_{i+1}}{2^{m-i-1}}$$

$$\leq \frac{A_i}{2^{m-i-1}} + \frac{1}{2^{m-i-1}} \frac{C}{m}$$

where the first equality is Statement 2 of Lemma A.5, the first inequality is Statement 1 of A.2, and the last inequality follows from the fact that $\delta_i^{\alpha+1} = 1$, $\delta_{i+1}^{\alpha+1} = 0$, and $\delta_{i+2}^{\alpha+1} = 1$ implies that $\frac{H_i - H_{i+1}}{2^{m-i-1}} \leq \frac{1}{2^{m-i-1}} \frac{C}{m}$. ∎

**Lemma A.9.** *For any cycle $\alpha$ and any index $1 \leq i < m-2$, if $\delta_i^{\alpha+1} = 0$, $\delta_{i+1}^{\alpha+1} = 1$, and $\delta_{i+2}^{\alpha+1} = 0$, then:*

$$(\Psi'^{\alpha+1} - \Psi'^{\alpha})_{i+1,i+1} - \sum_{k=i+1}^{i+1} \frac{\omega_k}{2^{m-k}} + \frac{A_i}{2^{m-i-1}} \leq \frac{A_i}{2^{m-i-1}} + \frac{1}{2^{m-i-1}} \frac{C}{m} \qquad (28)$$

*Proof.* Consider:

$$(\Psi'^{\alpha+1} - \Psi'^{\alpha})_{i+1,i+1} - \sum_{k=i+1}^{i+1} \frac{\omega_k}{2^{m-k}} + \frac{A_i}{2^{m-i-1}} = \frac{A_i}{2^{m-i-1}} + \frac{A_{i+1}}{2^{m-i-1}}$$

$$\leq \frac{A_i + H_{i+1}^{\alpha+1} - H_{i+2}^{\alpha}}{2^{m-i-1}}$$

$$\leq \frac{A_i}{2^{m-i-1}} + \frac{1}{2^{m-i-1}} \frac{C}{m}$$

where the first equality is Statement 1 of Lemma A.5, the first inequality is Statement 2 of A.2, and the last inequality follows from the fact that $\delta_i^{\alpha+1} = 0$, $\delta_{i+1}^{\alpha+1} = 1$, and $\delta_{i+2}^{\alpha+1} = 0$ implies that $\frac{H_{i+1}^{\alpha+1} - H_{i+2}^{\alpha}}{2^{m-i-1}} \leq \frac{1}{2^{m-i-1}} \frac{C}{m}$. ∎

**Claim A.10.** *For any cycle $\alpha$, we have:*

$$Z^{\alpha} + (\Psi'^{\alpha+1} - \Psi'^{\alpha})_{m,m} \leq A_{m-1}^{\alpha} \tag{29}$$

*Proof.* Since $(H_m^{\alpha+1} - (m-m)\frac{C}{m}) = H_m^{\alpha+1} \geq 0$, we have that the second term of $\min(0, H_m^{\alpha+1} - (m-m)\frac{C}{m})$ always dominates, and hence for all cycles, $\delta_m^{\alpha+1} = 1$. Therefore, applying Claim A.5 (for $i = m$):

$$(\Psi'^{\alpha+1} - \Psi'^{\alpha})_{m,m} = A_{m-1}^{\alpha} - A_m^{\alpha} + \omega_m^{\alpha}$$
$$\leq A_{m-1}^{\alpha} - A_m^{\alpha}$$
$$= A_{m-1}^{\alpha} - Z^{\alpha} \tag{30}$$

where the inequality follows since $\omega_i^{\alpha} \leq 0$ for all cycles $\alpha$ and nodes $i$, and the last equality is because $N_m$ is the node that will be connected to the Receiver in the last $C$ rounds of $\alpha$, so by definition $A_m^{\alpha} = Z^{\alpha}$. ∎

We are now ready to prove the main result of this section, namely that (11) is satisfied for all cycles $\alpha$:

**Theorem A.11.** *For all cycles $\alpha$, the following is always true:*

$$Z'^{\alpha} + (\Psi'^{\alpha+1} - \Psi'^{\alpha}) \leq 7\frac{C}{m},$$

*Proof.* Fix cycle $\alpha$, and consider the string of bits $\{\delta_i^{\alpha+1}\}_{i=1}^{m}$:

$$(\delta_1^{\alpha+1}, \delta_2^{\alpha+1}, \ldots, \delta_{m-1}^{\alpha+1}, \delta_m^{\alpha+1}) \tag{31}$$

By Claim A.10, we have:

$$Z^{\alpha} + \Psi'^{\alpha+1} - \Psi'^{\alpha} = Z^{\alpha} + (\Psi'^{\alpha+1} - \Psi'^{\alpha})_{1,m} \leq (\Psi'^{\alpha+1} - \Psi'^{\alpha})_{1,m-1} + A_{m-1}^{\alpha} \tag{32}$$

We now use Lemmas A.6, A.7, A.8, and A.9 on the appropriate indices (based on the form of $\{\delta_i^{\alpha+1}\}$), which yields:[19]

1. For the smallest index $i$ such that $\delta_i^{\alpha+1} = 1$, we have leading term:

$$\frac{A_i}{2^{m-1}} \tag{33}$$

---

[19]We combine these lemmas by starting at the far right index $i = m-1$, and working our way down through smaller indices by using the appropriate lemma. Notice that the first term on the RHS of the inequality of each lemma is exactly the term needed on the LHS of the next lemma.

2. For any indices $i < j$ falling under Lemma A.6, we have contributions:

$$\frac{j-i+1}{2^{m-i+1}}(A_{i-1} + H_i) + \sum_{k=i+1}^{j-1} \frac{(j-k+1)(m-i)}{2^{m-k+2}} \quad (34)$$

3. For any indices $i < j$ falling under Lemma A.7, we have contribution:

$$\sum_{k=i}^{j} \frac{m-i}{2^{m-k+1}} \quad (35)$$

4. For any indices $i+1$ falling under Lemma A.8 or A.9, we have contribution:

$$\frac{1}{2^{m-i-1}} \frac{C}{m} \quad (36)$$

Notice that in terms of the contributions from (34), $(A_{i-1} + H_i) \leq \frac{(m-i-1)C}{m}$ by Statement 2 of Lemma A.2 together with the fact that $\delta_{i-1}^{\alpha+1} = 0$ implies $H_{i-1}^{\alpha+1} < \frac{(m-i-1)C}{m}$. The theorem now follows immediately from the facts:

1. For any $1 \leq i < j < \infty$, $\sum_{k=i}^{j} \frac{1}{2^n} \leq \sum_{k=1}^{\infty} \frac{1}{2^n} = 1$

2. For any $1 \leq i < j < \infty$, $\sum_{k=i}^{j} \frac{n}{2^n} \leq \sum_{k=1}^{\infty} \frac{n}{2^n} = 2$

3. For any $1 \leq i < j < \infty$, $\sum_{k=i}^{j} \frac{n(n-1)}{2^n} \leq \sum_{k=1}^{\infty} \frac{n(n-1)}{2^n} = 4$

■

The remainder of the proof that the optimal competitive ratio is $1/n$ was presented in Section 3.

## B  Rigorous Proof of Competitive Ratio of Slide

The high-level ideas of the proof of Theorem 2.2 were sketched in Section 4.2, and we encourage the reader to re-read that section before proceeding here. In this Section, we begin by providing in Section B.1 a deeper explanation of the proof than was provided in Section 4.2, but still does not go into the details of the proofs. Then in Sections B.2-B.5 we rigorously prove all the lemmas and theorems.

### B.1  Motivation and Definitions

In what follows, unless stated otherwise, all notation is as defined in Section 4.2. Recall from Section 4.2 that we wish to construct two potential functions. The first one, denoted by $\varphi_{p'}$, will be associated to every packet $p' \in Z_2^{\mathcal{P}'}$. However, $\varphi_{p'}$ will not be exactly as defined in Section 4.2, so we provide now the motivation to explain how $\varphi_{p'}$ is actually defined, and why we need to slightly change what it represents.

Our first attempt employed in Section 4.2 was to define $\varphi_{p'}$ to be the height, *with respect to* $\mathcal{P}$, of the node in which $p'$ was currently being stored. **We state once-and-for-all that when referencing the *height* of a node, we will mean its height with respect to the Slide protocol** $\mathcal{P}$. As noted in Section 4.2, if we define $\varphi_{p'}$ this way, then for every $p' \in Z_2^{\mathcal{P}'}$, $\varphi_{p'}$ will be initially set to $C$ (when $\mathcal{P}'$ first inserts $p'$), and $\varphi_{p'}$ will be zero when $p'$ is delivered to the Receiver. Thus, there is a net change of $-C$ to $\varphi_{p'}$ from the time of insertion by the Sender to the time of reception by the Receiver. The goal was then to define a second overall network potential function $\Phi$, which increases by $C$ every time $\mathcal{P}$ transfers a packet, and such that any time $\varphi_{p'}$ changes for any $p' \in Z_2^{\mathcal{P}'}$, the cumulative changes of $\sum_{p' \in Z_2^{\mathcal{P}'}} \varphi_{p'}$ will be mimicked by $\Phi$. Since $\Phi$ increases by $C$ when there is a packet transfer in $\mathcal{P}$, one (good) way to think of this approach is that for each drop in $\varphi_{p'}$, we would like to find a packet transfer in $\mathcal{P}$ that can be "charged," i.e. this packet transfer "allowed" $\varphi_{p'}$ to decrease.

Unfortunately, with the simplistic definition of $\varphi_{p'}$ equal to the height of the node it is currently stored in, we encounter a problem. To clarify the problem, as well as to set notation, at the very beginning of each round $x$, we will label the *internal* nodes (i.e. not the Sender or Receiver) as: $\{N_1^x, N_2^x, \ldots, N_{n-2}^x\}$, where the labeling respects heights, so that at the start of the round $x$, $N_{i+1}^x$ is storing at least as many packets as $N_i^x$ (ties are broken arbitrarily). Letting $H_i^x$ denote the height of $N_i^x$ at the start of $x$ (i.e. the number of packets $N_i^x$ is storing with respect to $\mathcal{P}$), we may restate the criterion for labeling nodes at the start of each round by writing: $H_1^x \leq H_2^x \leq \cdots \leq H_{n-2}^x$. Note that nodes may change labels from one round to the next, i.e. we may have $N_i^x \neq N_i^{x+1}$. When the round is unimportant, we will suppress the superscript $x$. Let $S$ denote the Sender and $R$ denote the Receiver.

We may now explain why the simplistic definition of $\varphi_{p'}$ above will not be adequate. Define $Q := \frac{C-n}{n}$, and consider the following two scenarios that may be present at the start of some round $x$:

**Scenario 1:**  $H_{n-2} = C$   $H_{n-3} = C$   $\ldots$   $H_3 = C$   $H_2 = C$   $H_1 = (n-3)Q$
**Scenario 2:**  $H_{n-2} = (n-3)Q$   $H_{n-3} = (n-4)Q$   $\ldots$   $H_3 = 2Q$   $H_2 = Q$   $H_1 = 0$

In Scenario 1, consider a packet $p' \in Z_2^{\mathcal{P}'}$ that begins round $x$ in node $N_1$, so that $\varphi_{p'} = (n-3)Q$. Notice that if the adversary honors the edge $E(N_1, R)$, the Slide protocol will transfer a packet to the Receiver (Rules 2 and 3a of Section 4.1). Now by definition of being in the set $Z_2^{\mathcal{P}'}$, in order for $p'$ to be delivered to the Receiver via node[20] $N_1$, node $N_1$ must have height zero when the adversary honors edge $E(N_1, R)$. Therefore, there must be exactly $(n-3)Q$ transfers in $\mathcal{P}$ (to drain $N_1$) before $p'$ can be delivered to $R$ via $N_1$. Thus, loosely speaking, we can "charge" the resulting drop in $\varphi_{p'}$ from $(n-3)Q$ to 0 to these $(n-3)Q$ transfers in $\mathcal{P}$.

Now instead imagine we are in Scenario 2, and again fix a packet $p' \in Z_2^{\mathcal{P}'}$ such that $\varphi_{p'} = (n-3)Q$ at the start of round $x$, so $p' \in N_{n-2}$. In this case, notice that $p'$ has a way to reach $R$ without *any* packets being transferred in $\mathcal{P}$. In particular, the adversary could honor edge $E(N_{n-2}, N_{n-3})$ in round $x$, and then $E(N_{n-3}, N_{n-4})$ in round $x+1$, and so forth. Since the difference in heights between adjacent nodes is less than $C/n$, the Slide protocol will not transfer any packets during these rounds. Meanwhile, protocol $\mathcal{P}'$ may dictate that $p'$ is transferred each of these rounds, all the way to the Receiver. Thus, in this scenario, $\varphi_{p'}$ was able to decrease from $(n-3)Q$ to zero *without any packets being transferred in $\mathcal{P}$*. Because we are trying to associate drops in $\varphi_{p'}$ to packet transfers in $\mathcal{P}$, this is problematic.

Notice that the problem in Scenario 2 is that there exists a "bridge" between $N_{n-2}$ and $R$. That is, even though $N_{n-2}$ has a relatively large height, there is still a way for packets $p' \in Z_2^{\mathcal{P}'}$ that are in $N_{n-2}$ to reach $R$ without $\mathcal{P}$ being able to transfer any packets. In contrast, in Scenario 1, $p' \in N_1$ will also have $\varphi_{p'} = (n-3)Q$, but now there must be $(n-3)Q$ transfers in $\mathcal{P}$ before $p'$ can reach $R$ (again, since $p' \in Z_2^{\mathcal{P}'}$ requires that $p'$ is never transferred at the same time as a packet in $\mathcal{P}$). In summary, one might say that even though node $N_1$ in Scenario 1 has the same height as node $N_{n-2}$ from Scenario 2, these two nodes have different "effectual" heights.

Considering the above two Scenarios, we were encouraged to modify our definition of $\varphi_{p'}$ as follows:

- For node $N_i$, define the node's effectual height:[21] $\widetilde{H}_i := \max(0, H_i - (i-1)\frac{C}{n})$

---
[20] Of course there is no reason to assume that $p'$ must be transferred to $R$ via $N_1$, but for the sake of the example, we imagine this is the case.
[21] The "maximum" is added to prevent the effectual height of a node from being negative.

- For any $p' \in Z_2^{\mathcal{P}'}$ that is currently in $N_i$, define its potential: $\varphi_{p'} := \widetilde{H}_i$

This is *almost* the actual definition we eventually make for $\varphi$, but we will need to first "smooth-out" this definition. To motivate the need to smooth the definition, consider the following events, which represent the only ways that $\varphi_{p'}$ can change (based on the new definition of $\varphi_{p'}$):

Case 1. $p'$ is transferred from $N_i$ to $N_j$ in some round $E(N_i, N_j)$

Case 2. $p' \in N_i$ when $N_i$ changes height due to a packet transfer in $\mathcal{P}$, but this packet transfer does *not* cause a re-indexing of nodes

Case 3. $p'$ is in some node $N_i$ when a packet transfer in $\mathcal{P}$ causes $N_i$ to change index to $N_j$ (i.e. this node moves from the $i^{th}$ fullest node to the $j^{th}$ fullest node)

Since we are only concerned with $p' \in Z_2^{\mathcal{P}'}$, we note that whenever $\varphi_{p'}$ changes as by 1) above, necessarily $\mathcal{P}$ did not transfer a packet this round. In particular, this means that $|H_i - H_j| < C/n$. In order to control changes to $\varphi_{p'}$ that are a result of Case 1, we would therefore like for $\widetilde{H}_i \approx \widetilde{H}_j$ whenever $H_i \approx H_j$. Although the definition of effectual height $\widetilde{H}_i$ above almost captures this, there is necessarily a "jump" of $C/n$ between the values $\widetilde{H}_i$ and $\widetilde{H}_j$. This is one of the reasons we will want to "smooth-out" the definition of $\varphi_{p'}$.

Changes to $\varphi_{p'}$ that come from Case 2 above are okay, since in such cases $\varphi_{p'}$ will change by one, and this can be "charged" to the fact that there has been a packet transfer in $\mathcal{P}$. Lastly, notice that $\varphi_{p'}$ can only change as in Case 3 above if there are two nodes at the outset of some round $x$, $N_i$ and $N_{i+1}$, such that a packet transfer during round $x$ causes them to switch places (e.g. before the transfer, $H_i = H_{i+1}$, and then $N_i$ receives a packet in round $x$). Because there has been a packet transfer in $\mathcal{P}$, we can "charge" some of the changes in $\varphi_{p'}$ to this packet transfer, but again the fact that there will be a "jump" of $C/n$ to changes in $\varphi$ will encourage a "smoothing" of the definition of $\varphi$.

This leads to the notion of a family of nodes. In particular, we will partition the internal nodes into families. Intuitively, two nodes will be in the same family if they are relatively close to each other in height (or more generally, if there is a "bridge" connecting them, as in Scenario 2 above). Then within each family, we will distribute the cumulative effectual height of the nodes in that family evenly among all nodes in the family. Formally, for a family of nodes[22] $\mathcal{F} = \{N_i, N_{i+1}, \ldots, N_j\}$, define the cumulative effectual height $H_{\mathcal{F}}$ of the family $\mathcal{F}$ by:

$$\widetilde{H}_{\mathcal{F}} := \sum_{k=i}^{j} \widetilde{H}_k = \sum_{k=i}^{j} \max\left(0, H_k - (k-1)\frac{C}{n}\right)$$

For any $p' \in Z_2^{\mathcal{P}'}$ such that $p'$ is currently in some node of family $\mathcal{F}$, we will define $\varphi_{p'}$ to be the *average* effectual height of the family, i.e.:

$$\varphi_{p'} := \frac{\widetilde{H}_{\mathcal{F}}}{|\mathcal{F}|}$$

Of course, $\widetilde{H}_{\mathcal{F}}$ may not divide evenly among the nodes in the family $\mathcal{F}$, and then to force $\varphi_{p'} \in \mathbb{N}$, we will distribute the excess weight (the remainder) to the nodes with higher indices. Based on this definition of $\varphi_{p'}$, note that if $p'$ transfers between two nodes of the same family, $\varphi_{p'}$ can change by at most one.

We re-visit the three ways $\varphi_{p'}$ may change, explaining in each case how we can find a packet transfer in $\mathcal{P}$ to "charge" for the change in $\varphi_{p'}$. In terms of changes to $\varphi_{p'}$ resulting from Case 1 above, we recall that necessarily $|H_i - H_j| < C/n$. We show in Lemma B.12 that anytime $|H_i - H_j| < C/n$, $N_i$ and $N_j$ are necessarily in the same family, in which case our definition of $\varphi$ now guarantees that $\varphi_{p'}$ can change by at most one when $p'$ is transferred between nodes. Changes to $\varphi_{p'}$ due to Case 2 will be at most one (since the cumulative effectual height of the family will change by at most one, and this change will be distributed among nodes in the family), and we can "charge" such changes to the packet transfer in $\mathcal{P}$ that caused Case 2 to occur. Finally, for Case 3, if $p' \in N_i$ when $N_i$'s index changes but $N_i$ remains in the same family, than since $\varphi$ is distributed evenly among nodes in the family, the change in index will be irrelevant (i.e. this will not cause $\varphi_{p'}$ to change). On the other hand, we will show that whenever a node $N_i$ switches *families* as a result of a packet transfer in $\mathcal{P}$, the average effectual height of its new family will differ by at most one from the average effectual height of its old family. Thus, in this case the change in $\varphi_{p'}$ is also bounded by one, and we can "charge" this change to the packet transfer that caused families to re-align.

---

[22]We will show in the next section that nodes within the same family will always have adjacent indices.

Defining how to partition nodes into families so that the families behave the way we want (e.g. so that: 1) nodes with height within $C/n$ of each other are in the same family; 2) Families can only re-align during a round in which $\mathcal{P}$ transfers a packet; and 3) When families re-align, the average effectual height of any node before and after the re-alignment differs by at most one) requires a little thought, and it is done precisely in the following section. Once we have the formal definition of a family, we would like to formalize the notion of "charging a change in $\varphi_{p'}$ to a packet transfer in $\mathcal{P}$." Namely, as mentioned in Section 4.2, we define a second network potential $\Phi$ that will increase by $C$ every time there is a packet transfer in $\mathcal{P}$, and that will also mirror the cumulative changes of $\varphi_{p'}$ for each $p' \in Z_2^{\mathcal{P}'}$. In order to prove $\Phi$ is always positive, we will distribute the total network potential between the families:

$$\Phi = \Phi_{\mathcal{F}_1} + \cdots + \Phi_{\mathcal{F}_l} \tag{37}$$

and then show in Lemma B.17 that within each family $\mathcal{F}$:

$$\Phi_{\mathcal{F}} \geq 0. \tag{38}$$

The careful definition of families and the precise definition of the potential $\varphi$ and the network potential $\Phi$ is presented below in Section B.2. The main lemma and proof of the fact that at all times $\Phi \geq 0$ can be found in Section B.5.

## B.2 Formal Definition of "Family" and Potential of a Packet ($\varphi_{p'}$)

We begin by defining formally the notion of a family introduced in the previous section. Note that families will in general re-align during a round when there is a packet transfer in $\mathcal{P}$, so we use the notation $\mathcal{F}^x$ to denote some family $\mathcal{F}$ that was in existence at the start of round $x$. Recall that at the start of each round $x$, the internal nodes are indexed according to their heights with respect to $\mathcal{P}$: $\{N_1, N_2, \ldots, N_{n-2}\}$, so that $H_i \leq H_j$ if $i < j$ (ties are broken arbitrarily). Also recall from the previous section the definition of the effectual height $\widetilde{H}_i$ of node $N_i$:

$$\widetilde{H}_i := \max\left(0, H_i - (i-1)\frac{C}{n}\right) \tag{39}$$

At the start of each round, we will partition the internal nodes into families inductively (starting from the emptiest nodes), so that the average effectual height of each family is minimized. In particular:

**Definition B.1.** At the start of round $x$, internal nodes will be partitioned into families $\{\mathcal{F}_i^x\}$ as follows. Starting at $i = 1$ and $k_0 = 0$:

**F1** *Find index $k_{i-1} < k_i \leq n-2$ such that the following quantity is **minimal**:*

$$\frac{\sum_{j=k_{(i-1)}+1}^{k_i} \widetilde{H}_j}{(k_i - k_{i-1})} \tag{40}$$

*In case there are multiple values for $k_i$ that achieve the same minimum, define $k_i$ to be the **largest** of all possibilities. Then define*[23] *family $\mathcal{F}_i^x := \{N_{k_{(i-1)}+1}^x, \ldots, N_{k_i}^x\}$.*

**F2** *Set $i = i + 1$ and repeat Step F1 until all internal nodes are in some family.*

**F3** *The Sender and Receiver will form their own, separate, families. Denote the Sender's family by $\mathcal{F}_n$ and the Receiver's family by $\mathcal{F}_0$.*[24]

**Definition B.2.** The cumulative effectual height $\widetilde{H}_{\mathcal{F}}$ of a family $\mathcal{F}$ is the sum of the effectual heights of each of the nodes in the family. The average effectual height $\langle \widetilde{H}_{\mathcal{F}} \rangle$ of a family is the cumulative effectual height divided by the size of the family. Succinctly, if $\mathcal{F} := \{N_i, N_{i+1}, \ldots, N_j\}$:

$$\widetilde{H}_{\mathcal{F}} := \sum_{k=i}^{j} \widetilde{H}_k \quad \text{and} \quad \langle \widetilde{H}_{\mathcal{F}} \rangle := \frac{\widetilde{H}_{\mathcal{F}}}{|\mathcal{F}|} = \frac{\sum_{k=i}^{j} \widetilde{H}_k}{j - i + 1} \tag{41}$$

Notice that by construction (see Rules F1 and F2), families are created so that the average effectual height of (the lowest indexed) families is minimized.

With the formal definition of families in hand, we are ready to formally define the first kind of potential, $\varphi$. Recall that this potential will be associated to packets $p' \in Z_2^{\mathcal{P}'}$, and if $p' \in N_i \in \mathcal{F}$ at the start of some round, then $\varphi_{p'}$ will (roughly) represent the average effectual height $\langle \widetilde{H}_{\mathcal{F}} \rangle$. More precisely, we will ascribe to each node $N_i \in \mathcal{F}$ a potential $\varphi_i$ equal to the average effectual height, except that the potential for some nodes in the family will be one bigger to account for the case that $\frac{\widetilde{H}_{\mathcal{F}}}{|\mathcal{F}|} \notin \mathbb{Z}$. Formally:

**Definition B.3.** Let $\mathcal{F} = \{N_i, N_{i+1}, \ldots, N_j\}$. Then the potential $\varphi_k$ of a node $N_k \in \mathcal{F}$ will be either $\langle \widetilde{H}_{\mathcal{F}} \rangle$ or $\langle \widetilde{H}_{\mathcal{F}} \rangle + 1$. More precisely, writing:

$$\widetilde{H}_{\mathcal{F}} = \lfloor \langle \widetilde{H}_{\mathcal{F}} \rangle \rfloor * |\mathcal{F}| + r \tag{42}$$

Then define subsets of $\mathcal{F}$:

$$\mathcal{F}^- := \{N_i, N_{i+1}, \ldots, N_{j-r}\} \quad \text{and} \quad \mathcal{F}^+ := \{N_{j-r+1}, \ldots, N_j\} \tag{43}$$

Then for nodes $N_k \in \mathcal{F}^+$, define $\varphi_k = \lfloor \langle \widetilde{H}_{\mathcal{F}} \rangle \rfloor + 1$. For nodes $N_k \in \mathcal{F}^-$, define $\varphi_k = \lfloor \langle \widetilde{H}_{\mathcal{F}} \rangle \rfloor$. Finally, if $p' \in Z_2^{\mathcal{P}'}$ and $p'$ is currently being stored in $N_k$, then define the potential $\varphi_{p'}$ to be the potential of $N_k$, i.e. $\varphi_{p'} := \varphi_k$.

One immediate consequence of the above definition that we will need later is:

**Lemma B.4.** *At the beginning of any round $x$ and for any family $\mathcal{F}^x$, the sum of the potentials for the nodes in $\mathcal{F}$ equals the cumulative effectual height of the family:*

$$\sum_{N \in \mathcal{F}} \varphi_N = \widetilde{H}_{\mathcal{F}} \tag{44}$$

**Definition B.5.** The network potential $\Phi$ is an integer satisfying the following properties:

1. $\Phi$ begins the protocol equal to zero.

2. $\Phi$ increases by $4C$ every time a packet is transferred in protocol $\mathcal{P}$

3. For any packet $p' \in Z_2^{\mathcal{P}'}$, any time $\varphi_{p'}$ changes, $\Phi$ changes by the same amount.

---

[23] When the round $x$ is unimportant, we will suppress the superscript in our notation.

[24] The only reason we place the Sender and Receiver in a family at all is to make the terminology easier in the lemmas that follow. In particular, the notation we use for the Sender's family ensures that it will have a higher index than all other nodes (there will be a gap between the index of the largest indexed family of internal nodes and the Sender's family, which is unimportant), and conversely the Receiver's family will have a smaller index than all other nodes.

## B.3 Preliminary Lemmas

In this section, we state and prove the basic properties that follow from the definitions of the previous section.

**Lemma B.6.** *At all times, all families consist of nodes with adjacent indices. In particular, if at the start of any round $x$ there are $l$ families, then there exist indices $k_1 < k_2 < \cdots < k_{l-1}$ such that:*

$$\mathcal{F}_1 = \{N_1, \ldots, N_{k_1}\}, \quad \mathcal{F}_2 = \{N_{k_1+1}, \ldots, N_{k_2}\}, \ldots, \mathcal{F}_l = \{N_{k_{l-1}+1}, \ldots, N_{n-2}\} \quad (45)$$

*Proof.* This follows immediately from the rules regarding the construction of families (see F1 and F2 in the previous section). ∎

**Lemma B.7.** *Fix some round $x$ and some pair of nodes $N_i^x$ and $N_j^x$ for $i < j$. Then:*

1. *If $H_i^x \geq H_j^x - C/n$, then $\widetilde{H}_i^x \geq \widetilde{H}_j^x$.*

2. *If $H_i^x < H_j^x - (j-i)C/n$ and $\widetilde{H}_j > 0$, then $\widetilde{H}_i^x < \widetilde{H}_j^x$.*

*Proof.* Consider the following string of inequalities:

$$\begin{aligned}
\widetilde{H}_i - \widetilde{H}_j &= \max(0, H_i - (i-1)C/n) - \max(0, H_j - (j-1)C/n) \\
&\geq \max(0, H_i - (i-1)C/n) - \max(0, (H_i + C/n) - (j-1)C/n) \\
&\geq \max(0, H_i - (i-1)C/n) - \max(0, (H_i + C/n) - ((i+1) - 1)C/n) \\
&= \max(0, H_i - (i-1)C/n) - \max(0, (H_i - (i-1)C/n) \\
&= 0
\end{aligned}$$

This proves Statement 1. For Statement 2, if $\widetilde{H}_i = 0$, then it is immediate. Otherwise, consider the inequalities:

$$\begin{aligned}
\widetilde{H}_j - \widetilde{H}_i &= H_j - (j-1)C/n - (H_i - (i-1)C/n) \\
&= H_j - H_i + ((i-1) - (j-1))C/n \\
&> (j-i)C/n + (i-j)C/n \\
&= 0
\end{aligned}$$
∎

We state a trivial observation regarding fractions of positive numbers that will be useful in proving the lemmas below.

**Observation 2.** *For any positive numbers $a, b, c, d \in \mathbb{N}$:*

1. $\frac{a}{b} < \frac{c}{d} \Rightarrow \frac{a}{b} < \frac{a+c}{b+d} < \frac{c}{d}$

2. $\frac{a}{b} = \frac{c}{d} \Rightarrow \frac{a}{b} = \frac{a+c}{b+d} = \frac{c}{d}$

**Lemma B.8.** *Let $x$ be any round, and suppose that at the outset of the round there is some family $\mathcal{F}_\alpha^x = \{N_i, N_{i+1}, \ldots, N_j\}$. Then the following statements are all true at the outset of round $x$:*

1) *For any $i \leq k < j$:* $\quad \dfrac{\sum_{m=i}^{k} \widetilde{H}_m}{k - i + 1} \geq \langle \widetilde{H}_{\mathcal{F}_\alpha} \rangle \geq \dfrac{\sum_{m=k+1}^{j} \widetilde{H}_m}{j - k}$

2) *For any $j < k \leq n - 2$:* $\quad \langle \widetilde{H}_{\mathcal{F}_\alpha} \rangle < \dfrac{\sum_{m=j+1}^{k} \widetilde{H}_m}{k - j}$

3) $\langle \widetilde{H}_{\mathcal{F}_\alpha} \rangle < \langle \widetilde{H}_{\mathcal{F}_{\alpha+1}} \rangle$

*Proof.* The fact that $\frac{\sum_{m=i}^{k} \widetilde{H}_m}{k-i+1} \geq \frac{\sum_{m=k+1}^{j} \widetilde{H}_m}{j-k}$ follows immediately from Observation 2 together with the rules regarding the construction of families (see Rule F1 from the previous section), and in particular the fact that indices are found by *minimizing* (40). Statement 1 now follows from Observation 2. Statement 2 also follows immediately from Rule F1 and Observation 2, and Statement 3 follows immediately from Statement 2. ∎

Statement 3 of Lemma B.8 can be immediately extended:

**Corollary B.9.** *Let $x$ be any round, and suppose that at the outset of the round there are $l$ families. Then:*
$$\langle \widetilde{H}_{\mathcal{F}_1} \rangle < \langle \widetilde{H}_{\mathcal{F}_2} \rangle < \cdots < \langle \widetilde{H}_{\mathcal{F}_l} \rangle$$

**Lemma B.10.** *Let $x$ be any round, and suppose that at the outset of the round there is some family $\mathcal{F}_\alpha^x = \{N_i, N_{i+1}, \ldots, N_j\}$. Then:*

$$\text{For any } 1 \leq k < i: \quad \frac{\sum_{m=k}^{i-1} \widetilde{H}_m}{i-k} < \langle \widetilde{H}_{\mathcal{F}_\alpha} \rangle \tag{46}$$

*Proof.* Since $k < i$, necessarily $N_k$ is in some family $\mathcal{F}_\beta$ with index $\beta < \alpha$. Then:

$$\frac{\sum_{m=k}^{i-1} \widetilde{H}_m}{i-k} \leq \langle \widetilde{H}_{\mathcal{F}_\beta} \rangle < \langle \widetilde{H}_{\mathcal{F}_{\beta+1}} \rangle < \ldots < \langle \widetilde{H}_{\mathcal{F}_{\alpha-1}} \rangle < \langle \widetilde{H}_{\mathcal{F}_\alpha} \rangle, \tag{47}$$

where the first inequality is from Statement 1 of Lemma B.8 and the other inequalities are from Corollary B.9. ∎

**Lemma B.11.** *If at the start of some round $x$ we have that $\widetilde{H}_{j+1}^x \leq \widetilde{H}_j^x$, then $N_j$ and $N_{j+1}$ are in the same family at the start of round $x$.*

*Proof.* Suppose for the sake of contradiction that they are not in the same family at the start of round $x$. Let $\mathcal{F}^x$ denote $N_j$'s family at the start of the round. By Lemma B.6 and the fact that $j$ and $j+1$ are adjacent indices, we must have that $\mathcal{F}^x = \{N_i, N_{i+1}, \ldots, N_j\}$ for some $i \leq j$. The key observation is that:

$$\frac{\widetilde{H}_{j+1}}{1} \leq \frac{\widetilde{H}_j}{1} \quad \Rightarrow \quad \frac{\widetilde{H}_{j+1}}{1} \leq \frac{\widetilde{H}_{j+1} + \widetilde{H}_j}{2} \leq \frac{\widetilde{H}_j}{1} \tag{48}$$

If $i = j$, then (48) contradicts Statement 2 of Lemma B.8 (set $k = j+1$). If $i < j$, then define:

$$A := \sum_{l=i}^{j-1} \widetilde{H}_l \quad \text{and} \quad B := j - i \tag{49}$$

Then by Lemma B.8:

$$\frac{\widetilde{H}_{j+1}}{1} \leq \frac{\widetilde{H}_j}{1} \leq \frac{A}{B} \quad \Rightarrow \quad \frac{\widetilde{H}_{j+1} + \widetilde{H}_j + A}{B+2} \leq \frac{\widetilde{H}_j + A}{B+1} = \langle \widetilde{H}_\mathcal{F} \rangle, \tag{50}$$

which contradicts Statement 1 of Lemma B.8. ∎

**Lemma B.12.** *If at the outset of any round $x$, we have that $|H_i^x - H_j^x| \leq C/n$ for any pair of nodes $N_i^x$ and $N_j^x$, then necessarily the nodes are in the same family at the start of round $x$.*

*Proof.* Suppose for the sake of contradiction that there exists some round $x$ and some pair of nodes $N_i^x$ and $N_j^x$ for which $|H_i^x - H_j^x| \leq C/n$, but these nodes are in different families. Since families consist of adjacent indices (Lemma B.6) and nodes are indexed according to their heights at the start of the round, we may assume without loss of generality that $i$ and $j$ are adjacent (i.e. that $j = i+1$). By definition of indexing, we must have $H_i \leq H_{i+1}$, which combined with the hypothesis of the lemma implies that $H_{i+1} - C/n \leq H_i$. But then $\widetilde{H}_i \geq \widetilde{H}_{i+1}$ by Lemma B.7, and then $N_i^x$ and $N_{i+1}^x$ in different families contradicts Lemma B.11. ∎

## B.4 Lemmas Regarding the Re-structuring of Families

In this section, we discuss all possible changes between how families are arranged at the beginning of one round and the next.

**Lemma B.13.** *Families can only re-align during rounds $E(N_a, N_b)$ during which there is a packet transfer in $\mathcal{P}$ from $N_a$ to $N_b$.*

*Proof.* This is immediate from the rules regarding constructing families, since the values of $\{\widetilde{H}_i\}$ (39) can only change if there is a packet transfer in $\mathcal{P}$, and thus the analysis in Rule F1 (40) will not change if there has been no packet transfer in $\mathcal{P}$. ∎

**Lemma B.14.** *Suppose that in some round $x = E(N_a, N_b)$, the Slide protocol transfers a packet from $N_a$ to $N_b$. Let $\mathcal{F}_\alpha := \{N_e, \ldots, N_a, \ldots, N_f\}$ denote $N_a$'s family at the start of round $x$ ($e \leq a \leq f$), and $\mathcal{F}_\beta := \{N_c, \ldots, N_b, \ldots, N_d\}$ denote $N_b$'s family[25] at the start of $x$ ($c \leq b \leq d$). The following describes all possible changes to the way families are organized between the start of round $x$ and the next round:*

<u>CASE 1: $\widetilde{H}_a$ AND $\widetilde{H}_b$ DO NOT CHANGE.</u> *Then the families at the start of round $x+1$ are identical the arrangement of families at the start of $x$.*

<u>CASE 2: $\widetilde{H}_a$ DOES NOT CHANGE, AND $\widetilde{H}_b$ INCREASES BY ONE.</u> *Then:*

(a) *Families $\mathcal{F}_\delta$ to the left of $\mathcal{F}_\beta$ (i.e. $\delta < \beta$) do not change*

(b) *For any node $N_m$ with $b \leq m \leq d$, $N_m$ will be in the same family as $N_b$ at the start of round $x+1$*

(c) *For any node $N_m$ with $d < m$, letting $\mathcal{F}_\mu^x$ denote $N_m$'s family at the start of round $x$, one of the following happens:*

   i. $\mathcal{F}_\mu^x$ *does not change*

   ii. *Every node in $\mathcal{F}_\mu^x$ is in the same family as $N_b$ at the start of $x+1$*

<u>CASE 3: $\widetilde{H}_a$ DECREASES BY ONE, AND $\widetilde{H}_b$ DOES NOT CHANGE.</u> *Then:*

(a) *Families $\mathcal{F}_\delta$ to the right of $\mathcal{F}_\alpha$ (i.e. $\delta > \alpha$) do not change*

(b) *For any node $N_m$ with $e \leq m \leq a$, $N_m$ will be in the same family as $N_a$ at the start of round $x+1$*

(c) *For any node $N_m$ with $m < e$, letting $\mathcal{F}_\mu^x$ denote $N_m$'s family at the start of round $x$, one of the following happens:*

   i. $\mathcal{F}_\mu^x$ *does not change*

   ii. *Every node in $\mathcal{F}_\mu^x$ is in the same family as $N_a$ at the start of $x+1$*

<u>CASE 4: $\widetilde{H}_a$ DECREASES BY ONE, AND $\widetilde{H}_b$ INCREASES BY ONE.</u> *Then:*

(a) *Families $\mathcal{F}_\delta$ to the right of $\mathcal{F}_\alpha$ (i.e. $\delta > \alpha$) and to the left of $\mathcal{F}_\beta$ (i.e. $\delta < \beta$) do not change*

(b) *For any node $N_m$ with $e \leq m \leq a$, $N_m$ will be in the same family as $N_a$ at the start of round $x+1$*

(c) *For any node $N_m$ with $b \leq m \leq d$, $N_m$ will be in the same family as $N_b$ at the start of round $x+1$*

---

[25]Note that necessarily $\beta \leq \alpha$, as if both $N_a$ and $N_b$ are internal nodes, then Rule 3 of the Slide protocol (together with the definition of how nodes are indexed) guarantees that $b < a$, and then $\beta \leq \alpha$ by Lemma B.6. If $N_a$ is the Sender and/or $N_b$ is the Receiver, then $\beta \leq \alpha$ comes from our choice to denote the Sender's family by $\mathcal{F}_n$ and the Receiver's family by $\mathcal{F}_0$ (see Rule F3 regarding the formation of families).

(d) For any node $N_m$ with $d < m < e$, letting $\mathcal{F}_\mu^x$ denote $N_m$'s family at the start of round $x$, one of the following happens:
   i. $\mathcal{F}_\mu^x$ does not change
   ii. Every node in $\mathcal{F}_\mu^x$ is in the same family as $N_a$ at the start of $x+1$
   iii. Every node in $\mathcal{F}_\mu^x$ is in the same family as $N_b$ at the start of $x+1$
   iv. Every node in $\mathcal{F}_\mu^x$ is in the same family as $N_a$ AND $N_b$ at the start of $x+1$

*Proof.* That the four cases stated in the lemma cover all possibilities is immediate from the definition of effective height $\widetilde{H}$ (see Definition (39)). Case 1 follows immediately from the rules F1-F2 for forming families (see Definition B.1) since the effective heights have not changed. We go through each of the other cases, and prove each Statement.

Suppose that we are in Case 2, so that $\widetilde{H}_a$ does not change, and $\widetilde{H}_b$ increases by one. For $\delta < \beta$, consider a family $\mathcal{F}_\delta := \{N_i, \ldots, N_j\}$, and for the sake of contradiction, suppose that $\mathcal{F}_\delta$ changes in some way from the start of round $x$ to the start of round $x+1$. Without loss of generality, we will suppose that $\delta < \beta$ is the *minimal* index for which $\mathcal{F}_\delta$ changes.

**Case A:** $\mathcal{F}_\delta$ *Splits.* In other words, $N_i$ and $N_j$ are *not* in the same family at the start of round $x+1$. Let $\mathcal{F}_\iota^{x+1} := \{N_i, \ldots, N_k\}$ denote $N_i$'s new family at the start of $x+1$, where $k < j$ by assumption.[26] Notice that for all $i \leq m \leq j$, the effective height $\widetilde{H}_m$ will not change between the start of $x$ and $x+1$ (since $j < b < a$). Therefore:

$$\frac{\sum_{l=k+1}^{j} \widetilde{H}_l}{j-k} \leq \frac{\sum_{l=i}^{k} \widetilde{H}_l}{k-i+1} = \langle \widetilde{H}_{\mathcal{F}_\iota^{x+1}} \rangle < \frac{\sum_{l=k+1}^{j} \widetilde{H}_l}{j-k}, \qquad (51)$$

where the first inequality is Statement 1 of Lemma B.8 and the last inequality is Statement 2 of Lemma B.8. Clearly (51) is impossible, yielding the desired contradiction.

**Case B:** $\mathcal{F}_\delta$ *Grows.* In other words, at the start of round $x+1$ there is some family $\mathcal{F}_\iota^{x+1} := \{N_i, \ldots, N_k\}$ for $k > j$. If $k < b$, then for all $i \leq m \leq k$, the effective height $\widetilde{H}_m$ will not change between the start of $x$ and $x+1$, so:

$$\frac{\sum_{l=i}^{j} \widetilde{H}_l}{j-i+1} < \frac{\sum_{l=j+1}^{k} \widetilde{H}_l}{k-j} \leq \frac{\sum_{l=i}^{j} \widetilde{H}_l}{j-i+1}, \qquad (52)$$

where the first inequality is Statement 2 of Lemma B.8 and the second inequality is Statement 1 of Lemma B.8. Clearly (52) is impossible, yielding the desired contradiction. On the other hand, if $k \geq b$, then for all $i \leq m \leq k$ and $m \neq b$, the effective height $\widetilde{H}_m$ will not change between the start of $x$ and $x+1$, but the effective height $\widetilde{H}_b$ increases by one from the start of $x$ and $x+1$. Therefore (using superscripts only when necessary to specify the round):

$$\frac{\sum_{l=i}^{j} \widetilde{H}_l}{j-i+1} < \frac{\sum_{l=j+1}^{k} \widetilde{H}_l^x}{k-j} < \frac{\sum_{l=j+1}^{k} \widetilde{H}_l^{x+1}}{k-j} \leq \frac{\sum_{l=i}^{j} \widetilde{H}_l}{j-i+1}, \qquad (53)$$

where the first inequality is Statement 2 of Lemma B.8 and the last inequality is Statement 1 of Lemma B.8. Clearly (53) is impossible, yielding the desired contradiction.

This proves Statement (a) of Case 2. For Statement (b), fix index $m \in [b, d]$ (Statement (b) is trivially true for $m = b$, so assume $b < m \leq d$). For the sake of contradiction, suppose that $N_m$ is *not* in the same family as $N_b$ at the start of $x+1$. Let $\mathcal{F}_\beta^{x+1} := \{N_i, \ldots, N_b, \ldots, N_j\}$ denote $N_b$'s new family at the start of $x+1$, so by assumption $j < m \leq d$, and also $c \leq i$ by Statement (a) of

---

[26] Necessarily $N_i$ is the smallest-indexed node in $\mathcal{F}_\iota$ by our choice of minimality for $\delta$.

Case 2. Notice that $\widetilde{H}_b^x + 1 = \widetilde{H}_b^{x+1}$, but that for all other $i \leq l \leq m$, $\widetilde{H}_l$ does not change from the start of $x$ and $x + 1$. If $i = c$ (using superscripts only when necessary to specify the round):

$$\frac{\sum_{l=j+1}^{d} \widetilde{H}_l}{d-j} \leq \frac{\sum_{l=c}^{j} \widetilde{H}_l^x}{j-c+1} < \frac{\sum_{l=c}^{j} \widetilde{H}_l^{x+1}}{j-c+1} < \frac{\sum_{l=j+1}^{d} \widetilde{H}_l}{d-j}, \tag{54}$$

where the first inequality is Statement 1 of Lemma B.8 and the last inequality is Statement 2 of Lemma B.8. Clearly (54) is impossible, yielding the desired contradiction. If on the other hand $c < i$, then (using superscripts only when necessary to specify the round):

$$\begin{aligned}
\frac{\sum_{l=j+1}^{d} \widetilde{H}_l}{d-j} &\leq \frac{\sum_{l=c}^{j} \widetilde{H}_l^x}{j-c+1} = \langle \widetilde{H}_{\mathcal{F}_\beta^x} \rangle \\
&\leq \frac{\sum_{l=c}^{i-1} \widetilde{H}_l^x}{i-c} \\
&< \frac{\sum_{l=c}^{i-1} \widetilde{H}_l^{x+1}}{i-c} \\
&< \langle \widetilde{H}_{\mathcal{F}_i^{x+1}} \rangle = \frac{\sum_{l=i}^{j} \widetilde{H}_l^{x+1}}{j-i+1} \\
&< \frac{\sum_{l=j+1}^{d} \widetilde{H}_l}{d-j},
\end{aligned} \tag{55}$$

where the first and second inequalities are both Statement 1 of Lemma B.8, the fourth inequality is Lemma B.10, and the last inequality is Statement 2 of Lemma B.8. Clearly (55) is impossible, yielding the desired contradiction.

This proves Statement (b) of Case 2. It remains to prove Statement (c). Fix some $m > d$, and let $\mathcal{F}_\mu^w = \{N_w, \ldots, N_m, \ldots, N_y\}$ denote $N_m$'s family at the start of $x$. We prove Statement (c) via the following two subclaims:

**Subclaim 1.** $\mathcal{F}_\mu$ does not Split. *In other words, $N_w$ and $N_y$ will be in the same family at the start of round $x + 1$.*

*Proof.* Suppose not. Let $\mathcal{F}_\omega^{x+1} = \{N_i, \ldots, N_w, \ldots, N_j\}$ denote $N_w$'s family at the start of round $x + 1$, so $c \leq i \leq w \leq j < y$ (where the first inequality is due to Statement (a)). Notice that for every $i \leq l \leq y$, the only possible effective height $\widetilde{H}_l$ that can possibly change in round $x$ is for $l = b$, in which case $\widetilde{H}_b^x + 1 = \widetilde{H}_b^{x+1}$. If $i = w$, then (using superscripts only when necessary to specify the round):

$$\frac{\sum_{l=w}^{j} \widetilde{H}_l}{j-w+1} < \frac{\sum_{l=j+1}^{y} \widetilde{H}_l}{y-j} \leq \frac{\sum_{l=w}^{j} \widetilde{H}_l}{j-w+1}, \tag{56}$$

where the first inequality is Statement 2 of Lemma B.8 and the second is Statement 1 of Lemma B.8. Clearly, (56) is impossible, yielding the desired contradiction. If on the other hand $i < w$, then (using superscripts only when necessary to specify the round):

$$\frac{\sum_{l=w}^{j} \widetilde{H}_l}{j-w+1} \leq \frac{\sum_{l=i}^{w-1} \widetilde{H}_l^{x+1} + \sum_{l=w}^{j} \widetilde{H}_l^x}{j-i+1} \leq \frac{\sum_{l=j+1}^{y} \widetilde{H}_l}{y-j} \leq \frac{\sum_{l=w}^{j} \widetilde{H}_l}{j-w+1}, \tag{57}$$

where the second inequality is Statement 2 of Lemma B.8, the third is Statement 1 of Lemma B.8, and the first comes from:

$$\frac{\sum_{l=w}^{j} \widetilde{H}_l}{j-w+1} \leq \frac{\sum_{l=i}^{w-1} \widetilde{H}_l^{x+1}}{w-i} \quad \Rightarrow \quad \frac{\sum_{l=w}^{j} \widetilde{H}_l}{j-w+1} \leq \frac{\sum_{l=i}^{w-1} \widetilde{H}_l^{x+1} + \sum_{l=w}^{j} \widetilde{H}_l^x}{j-i+1}, \tag{58}$$

where the first inequality is Statement 1 of Lemma B.8. Clearly, (57) is impossible, yielding the desired contradiction.

**Subclaim 2.** *If $\mathcal{F}_\mu$ gets larger, then necessarily $N_b$ will be in the same family as $N_w$ and $N_y$ at the start of round $x + 1$.*

*Proof.* Suppose not. Let $\mathcal{F}_\omega^{x+1} = \{N_i, \ldots, N_w, \ldots, N_j\}$ denote $N_w$'s family at the start of round $x + 1$, so $b < i \leq w \leq y \leq j$. Notice that for every $i \leq l \leq y$, since $b < i$, the effective height $\widetilde{H}_l$ does not change. If $i = w$, then since we are assuming $\mathcal{F}_\mu$ grows, we have $j > y$, and:

$$\frac{\sum_{l=w}^{y} \widetilde{H}_l}{y - w + 1} < \frac{\sum_{l=y+1}^{j} \widetilde{H}_l}{j - y} \leq \frac{\sum_{l=w}^{y} \widetilde{H}_l}{y - w + 1}, \tag{59}$$

where the first inequality is Statement 2 of Lemma B.8 and the second is Statement 1 of Lemma B.8. Clearly, (59) is impossible, yielding the desired contradiction. If on the other hand $i < w$ and $j > y$, then:

$$\frac{\sum_{l=i}^{w-1} \widetilde{H}_l}{w - i} < \frac{\sum_{l=w}^{y} \widetilde{H}_l}{y - w + 1} < \frac{\sum_{l=y+1}^{j} \widetilde{H}_l}{j - y}, \tag{60}$$

where the first inequality is from Lemma B.10, and the second is from Statement 1 of Lemma B.8. But then (60) implies:

$$\frac{\sum_{l=i}^{w-1} \widetilde{H}_l + \sum_{l=w}^{y} \widetilde{H}_l^x}{y - i + 1} < \frac{\sum_{l=y+1}^{j} \widetilde{H}_l}{j - y}, \tag{61}$$

which contradicts Statement 1 of Lemma B.8. Finally, if $i < w$ and $j = y$, then:

$$\frac{\sum_{l=i}^{w-1} \widetilde{H}_l}{w - i} < \frac{\sum_{l=w}^{y} \widetilde{H}_l}{y - w + 1}, \tag{62}$$

which contradicts Statement 1 of Lemma B.8.

Cases 3 and 4 follow analogous arguments. ∎

## B.5 Statement and Proof of Fact that Slide has Competitive Ratio $1/n$

**Lemma B.15.** *Suppose at the start of round $x$, there exists nodes $\{N_i^x, N_{i+1}^x, \ldots, N_j^x\}$ such that $H_i^x = \cdots = H_j^x$. Then under any permutation of the indices $\sigma : \{i, i+1, \ldots, j\} \to \{i, i+1, \ldots, j\}$, we have that:*

$$\sum_{k=i}^{j} \widetilde{H}_k^x = \sum_{k=i}^{j} \max(0, H_k^x - (k-1)C/n) = \sum_{k=i}^{j} \max(0, H_{\sigma(k)}^x - (k-1)C/n) \tag{63}$$

*In particular, the value for $\sum_{k=i}^{j} \widetilde{H}_k^x$ will not change if we re-index the nodes $\{N_i, \ldots, N_j\}$ in any arbitrary manner.*

*Proof.* This is immediate from the hypothesis that $H_i^x = H_{i+1}^x = \cdots = H_j^x$. ∎

**Lemma B.16.** *Suppose that in some round $x$, $N_a$ transfers a packet to $N_b$ in the Slide protocol. Let $\mathcal{F}_\beta$ denote $N_b$'s family and $\mathcal{F}_\alpha$ denote $N_a$'s family. Then either there is exactly one node $N_{b'} \in \mathcal{F}_\beta$ such that $\varphi_{b'}$ increases by one, or $\varphi_N$ does not change for every $N \in \mathcal{F}_\beta$. Similarly, either there is exactly one node $N_{a'} \in \mathcal{F}_\alpha$ such that $\varphi_{a'}$ decreases by one, or $\varphi_N$ does not change for every $N \in \mathcal{F}_\alpha$. No other node $N \in G$ will have $\varphi_N$ change as a result of this packet transfer.*

*Proof.* If $N_b$'s effectual height $\widetilde{H}_b$ does *not* increase as a result of the packet transfer (e.g. the '0' in the maximum statement of (39) dominates), then $\mathcal{F}_\beta$'s cumulative effectual height does not change, and as a result, the potential $\varphi$ of all nodes in $\mathcal{F}_\beta$ remains unchanged. If on the other hand $B$'s effectual height does increase, then this will raise the cumulative effectual height $\widetilde{H}_{\mathcal{F}_\beta}$ by one, and this will be absorbed by some node in $\mathcal{F}^-$. A similar argument works with respect to $N_a$ in $\mathcal{F}_\alpha$. The last statement of the lemma follows from Lemma B.4. ∎

We are now ready to prove the main lemma that will allow us to argue that the Slide protocol has competitive ratio $1/n$. To fix notation, for any internal node $N$, let $H_N^{\mathcal{P}'}$ denote the number of packets $p' \in Z_2^{\mathcal{P}'}$ that $N$ is currently storing. Recall the definition of $\Phi$ (see Definition B.5); we will distribute the overall potential $\Phi$ between all the families, and show that with the rules regarding changes in $\Phi$, the potential of a family is always positive. Namely:

**Lemma B.17.** *For every round $x$ and for all families $\mathcal{F}$ that are present at the start of $x$:*

$$\Phi \geq \sum_{\mathcal{F}} \max\left( \sum_{N \in \mathcal{F}^-} C - H_N^{\mathcal{P}'},\ \sum_{N \in \mathcal{F}^+} H_N^{\mathcal{P}'} \right) \geq 0 \tag{64}$$

*Proof.* We prove this based on induction on the round $x$. The lemma is clearly true at the outset of the protocol, when $\Phi = \Phi_{\mathcal{F}} = 0$, and all nodes are in the same family, since all nodes have height zero. Suppose that at the start of round $x = E(N_a, N_b)$, (64) is satisfied. We show that no matter what happens in round $x$, (64) will remain satisfied at the start of round $x + 1$.

Case 1: Neither $\mathcal{P}$ nor $\mathcal{P}'$ transfer a packet. In this case, families will not change (Lemma B.13), and no packets in $Z_2^{\mathcal{P}'}$ move, so there will be no changes to either side of (64).

Case 2: $\mathcal{P}'$ transfers a packet during $x$, but $\mathcal{P}$ does not. If the packet $p'$ transferred by $\mathcal{P}'$ is in $Z_1^{\mathcal{P}'}$, then neither side of (64) will change. So suppose $p' \in Z_2^{\mathcal{P}'}$. Note that in Case 1, $N_a$ and $N_b$ are in the same family, call it $\mathcal{F}$ (Since Slide does not transfer a packet, we have $|H_a - H_b| < C/n$, and see Lemma B.12).

- If $N_a$ and $N_b$ are in $\mathcal{F}^+$, then $\varphi_a = \varphi_b$, so $\varphi_{p'}$ does not change. In particular, neither side of (64) changes in this case. The same is true if $N_a$ and $N_b$ are both in $\mathcal{F}^-$

- If $N_a \in \mathcal{F}^+$ and $N_b \in \mathcal{F}^-$, then the change on the left-hand side of (64) is -1 (since $\Delta \varphi_{p'} = -1$), which matches the change on the right-hand side of (64) (since $H_b^{\mathcal{P}'}$ increases by one, and $H_a^{\mathcal{P}'}$ decreases by one). If instead $N_a \in \mathcal{F}^-$ and $N_b \in \mathcal{F}^+$, then similar reasoning shows that the change of both sides of (64) is $+1$.

Case 3: $\mathcal{P}$ transfers a packet from $N_a$ to $N_b$ in round $x$. Notice that this case is not concerned with whether or not $\mathcal{P}'$ also transfers a packet, as such a packet would necessarily be in $Z_1^{\mathcal{P}'}$ (by definition), and hence this packet movement in $\mathcal{P}'$ will not affect either side of (64). Also, without loss of generality $N_a$ is the sending node and $N_b$ is the receiving node. By Lemma B.14, there are 4 cases we must consider:

Case 3A: $\widetilde{H}_b$ and $\widetilde{H}_a$ do not change. Then by Lemma B.14, there will be no re-structuring of families between rounds $x$ and $x + 1$. Consequently, if $\mathcal{F}_\beta$ denotes $N_b$'s family and $\mathcal{F}_\alpha$ denotes $N_a$'s family (possible $\alpha = \beta$), then for all other families, (64) will remain valid. Also, $\varphi_N$ does not change for any $N \in \mathcal{F}_\beta$ (similarly for $N \in \mathcal{F}_\alpha$) since $\widetilde{H}_b$ and $\widetilde{H}_a$ do not change. Therefore, the right-hand side of (64) also will not change for $\mathcal{F}_\beta$ and $\mathcal{F}_\alpha$, and the only change in the left-hand side comes from the increase of $4C$ to $\Phi$ (see Rule 2 of Definition B.5), which can be divided arbitrarily among the families $\{\mathcal{F}\}$, and this will only help (64).

Case 3B: $\widetilde{H}_b$ increases by one, but $\widetilde{H}_a$ does not change. Let $\mathcal{F}_\beta = \{N_c, \ldots, N_b, \ldots, N_d\}$ for some $c \leq$ $b \leq d$. By Lemma B.14, there exist integers $r, s \geq 0$ and indices $\{k_1, \ldots, k_r\}$ and $\{l_1, \ldots, l_s\}$ such that $c \leq k_1 < \cdots < k_r \leq b \leq d < l_1 < \cdots < l_s$ and:

| Families at the start of $x$ | Families at the start of $x$ |
|---|---|
| $\mathcal{F}_\beta = \{N_c, \ldots, N_b, \ldots, N_d\}$ | $\widehat{\mathcal{F}}_\beta = \{N_c, \ldots, N_{k_1-1}\}$ |
| $\mathcal{F}_{\beta+1} = \{N_{d+1}, \ldots, N_{l_1-1}\}$ | $\widehat{\mathcal{F}}_{\beta+1} = \{N_{k_1}, \ldots, N_{k_2-1}\}$ |
| $\mathcal{F}_{\beta+2} = \{N_{l_1}, \ldots, N_{l_2-1}\}$ | $\widehat{\mathcal{F}}_{\beta+2} = \{N_{k_2}, \ldots, N_{k_3-1}\}$ |
| $\vdots$ | $\vdots$ |
| $\mathcal{F}_{\beta+s} = \{N_{l_{s-1}}, \ldots, N_{l_s-1}\}$ | $\widehat{\mathcal{F}}_{\beta+r-1} = \{N_{k_{r-1}}, \ldots, N_{k_r-1}\}$ |
| | $\widehat{\mathcal{F}}_{\beta+r} = \{N_{k_r}, \ldots, N_{l_s-1}\}$ |

and no other families change.

By Lemma B.16, there is only one node $N \in \mathcal{F}_\beta^-$ for which $\varphi_N$ increases by one as a result of the packet transfer. Although $\mathcal{F}_\beta$ will change in the manner described by the table above, by Lemma B.4, the *number* of nodes $N \in G$ with $\varphi_N = \lfloor \langle \widetilde{H}_{\mathcal{F}_\beta} \rangle \rfloor$ (respectively $\varphi_N = \lfloor \langle \widetilde{H}_{\mathcal{F}_\beta} \rangle \rfloor$) will not change (aside from the single node $N'$ for which $\varphi_{N'}$ increases by one, as guaranteed by Lemma B.16), although the specific nodes in $\mathcal{F}^+$ and $\mathcal{F}^-$ may vary. A simple computation ensures that the right-hand side of (64) changes in the exact same way as the left-hand side of (64) whenever any two nodes in $\mathcal{F}$ swap places (in $\mathcal{F}^+$ and $\mathcal{F}^-$). Therefore, we may assume without loss of generality that there is exactly one node $N' \in \mathcal{F}_\beta^-$ for which $\varphi_{N'}$ increases by one as a result of the packet transfer, and for all other nodes $N \in G$, $\varphi_N$ does not change between the start of $x$ and $x+1$.

For each $0 \leq i \leq r$ and $0 \leq j \leq s$, define the following quantities:

$$\begin{array}{l|l}
\text{Families at the start of } x & \text{Families at the start of } x \\
X_i = \sum_{N \in \widehat{\mathcal{F}}_{\beta+i}^-} (C - H_N^{\mathcal{P}'}) & \mathcal{X}_j = \sum_{N \in \mathcal{F}_{\beta+j}^-} (C - H_N^{\mathcal{P}'}) \\
Y_i = \sum_{N \in \widehat{\mathcal{F}}_{\beta+i}^+} H_N^{\mathcal{P}'} & \mathcal{Y}_j = \sum_{N \in \mathcal{F}_{\beta+j}^+} H_N^{\mathcal{P}'} \\
A_i = |\widehat{\mathcal{F}}_{\beta+i}^+| & \mathcal{A}_i = |\mathcal{F}_{\beta+i}^+| \\
B_i = |\widehat{\mathcal{F}}_{\beta+i}^-| & \mathcal{B}_i = |\mathcal{F}_{\beta+i}^-|
\end{array} \tag{65}$$

Also define $\mathcal{F}_* = \widehat{\mathcal{F}}_{\beta+r} \cup \mathcal{F}_\beta$, and:

$$\mu = \sum_{N \in \widehat{\mathcal{F}}_*^-} (C - H_N^{\mathcal{P}'}) \qquad \nu = \sum_{N \in \mathcal{F}_*^+} H_N^{\mathcal{P}'} \qquad \alpha = |\widehat{\mathcal{F}}_*^+| \quad \text{and} \quad \beta = |\mathcal{F}_*^-| \tag{66}$$

By the induction hypothesis, we have that at the start of round $x$:

$$\sum_{j=0}^{s} \Phi_{\mathcal{F}_{\beta+j}} \geq \sum_{j=0}^{s} \left( \frac{\mathcal{A}_j \mathcal{X}_j + \mathcal{B}_j \mathcal{Y}_j}{\mathcal{A}_j + \mathcal{B}_j} \right) \tag{67}$$

In addition to the above potential, we also have that $\Phi$ increases by $4C$ as a result of the packet transfer in Slide. Meanwhile, the goal is to show that at the start of round $x+1$:

$$\sum_{i=0}^{r} \Phi_{\widehat{\mathcal{F}}_{\beta+i}} \geq \sum_{i=0}^{r} \left( \frac{A_i X_i + B_i Y_i}{A_i + B_i} \right) \tag{68}$$

Putting all these facts together, we want to show that:

$$4C + \sum_{j=0}^{s} \left( \frac{\mathcal{A}_j \mathcal{X}_j + \mathcal{B}_j \mathcal{Y}_j}{\mathcal{A}_j + \mathcal{B}_j} \right) \geq \sum_{i=0}^{r} \left( \frac{A_i X_i + B_i Y_i}{A_i + B_i} \right) \tag{69}$$

We demonstrate in the remainder of the proof how to show (69) is satisfied.

First look at the term $i = r$ for the right-hand side of (69):

$$\frac{A_r X_r + B_r Y_r}{A_r + B_r} = \frac{(\alpha + 1 + \sum_{j=1}^{s} \mathcal{A}_j)(\mu + \sum_{j=1}^{s} \mathcal{X}_j - (C - H_{N'}^{\mathcal{P}'}))}{A_r + B_r}$$
$$+ \frac{(\beta - 1 + \sum_{j=1}^{s} \mathcal{B}_j)(\nu + H_{N'}^{\mathcal{P}'} + \sum_{j=1}^{s} \mathcal{Y}_j)}{A_r + B_r}$$
$$= \frac{\alpha + 1}{\alpha + \beta}(\mu - (C - H_{N'}^{\mathcal{P}'})) + \sum_{j=1}^{s} \mathcal{X}_j \frac{\mathcal{A}_j}{\mathcal{A}_j + \mathcal{B}_j} + \frac{\beta - 1}{\alpha + \beta}(\nu + H_{N'}^{\mathcal{P}'}) + \sum_{j=1}^{s} \mathcal{Y}_j \frac{\mathcal{B}_j}{\mathcal{A}_j + \mathcal{B}_j}$$
$$+ (\mathcal{Y}_1 - \mathcal{X}_1)\left(\frac{\alpha \sum \mathcal{B}_j - \beta \sum \mathcal{A}_j}{(\alpha + \beta)(A_r + B_r)}\right)$$
$$+ \cdots + (\mathcal{Y}_s - \mathcal{X}_s)\left(\frac{\mathcal{A}_s(\beta + \sum \mathcal{B}_j) - \mathcal{B}_s(\alpha + \sum \mathcal{A}_j)}{(\mathcal{A}_s + \mathcal{B}_s)(A_r + B_r)}\right)$$
$$< C + \frac{\alpha + 1}{\alpha + \beta}(\mu - (C - H_{N'}^{\mathcal{P}'})) + \sum_{j=1}^{s} \mathcal{X}_j \frac{\mathcal{A}_j}{\mathcal{A}_j + \mathcal{B}_j} + \frac{\beta - 1}{\alpha + \beta}(\nu + H_{N'}^{\mathcal{P}'}) + \sum_{j=1}^{s} \mathcal{Y}_j \frac{\mathcal{B}_j}{\mathcal{A}_j + \mathcal{B}_j}$$

We have used above that (by Lemmas B.8 and Corollary B.9):

$$\frac{\alpha}{\alpha + \beta} < \frac{\mathcal{A}_1}{\mathcal{A}_1 + \mathcal{B}_1} < \cdots < \frac{\mathcal{A}_s}{\mathcal{A}_s + \mathcal{B}_s} < \frac{1 + \alpha + \mathcal{A}_1 + \cdots + \mathcal{A}_s}{\alpha + \beta + \sum_{j=1}^{s}(\mathcal{A}_j + \mathcal{B}_j)} \tag{70}$$

Meanwhile, we look at the left-hand side of (69) for the $j = 0$ term:

$$\frac{\mathcal{A}_0 \mathcal{X}_0 + \mathcal{B}_0 \mathcal{Y}_0}{\mathcal{A}_0 + \mathcal{B}_0} = \frac{(\alpha + \sum_{i=0}^{r-1} A_i)(\mu + \sum_{i=0}^{r-1} X_i)}{\mathcal{A}_0 + \mathcal{B}_0}$$
$$+ \frac{(\beta + \sum_{i=0}^{r-1} B_i)(\nu + \sum_{i=0}^{r-1} Y_i)}{\mathcal{A}_0 + \mathcal{B}_0}$$
$$\geq \mu\left(\frac{\alpha}{\alpha + \beta}\right) + \nu\left(\frac{\beta}{\alpha + \beta}\right) - \frac{\mu + \sum_{i=0}^{r-1} X_i}{\mathcal{A}_0 + \mathcal{B}_0}$$
$$+ sum_{i=0}^{r-1} \frac{A_i X_i + B_i Y_i}{A_i + B_i}, \tag{71}$$

where we have used for the inequality above:

$$\frac{\mathcal{A}_0}{\mathcal{A}_0 + \mathcal{B}_0} < \frac{A_0}{A_0 + B_0} < \frac{A_1}{A_1 + B_1} < \cdots < \frac{A_{r-1}}{A_{r-1} + B_{r-1}} < \frac{1 + \alpha + \sum_{i=0}^{r-1} A_i}{\alpha + \beta + \sum_{i=0}^{r-1}(A_i + B_i)}, \tag{72}$$

with the inequalities following from Lemma B.8 and Corollary B.9. Putting this all together, we have that:

$$4C + \sum_{j=0}^{s}\left(\frac{\mathcal{A}_j \mathcal{X}_j + \mathcal{B}_j \mathcal{Y}_j}{\mathcal{A}_j + \mathcal{B}_j}\right) \geq \sum_{i=0}^{r}\left(\frac{A_i X_i + B_i Y_i}{A_i + B_i}\right)$$

which is (69).

The other cases are proven similarly. ∎

We state as an immediate consequence the lemma we needed in the discussion of Section 4:

**Lemma B.18.** *At all times:*

$$|Z_2^{\mathcal{P}'}| \leq 2nY^{\mathcal{P}} \leq 2n|Z^{\mathcal{P}}| + 2n^2 C \tag{73}$$

# C Competitive Analysis of the Slide+ Protocol

## C.1 Description of Slide+

Recall that we model an asynchronous network via a **scheduling adversary** that maintains a buffer of requests of the form $(u, v, p)$, which is a request from node $u$ to send packet $p$ to node $v$. The scheduling adversary proceeds in a sequence of honored edges (called **rounds**), whereby we will mean the following when we talk about an edge $E(u, v)$ being **honored** by the adversary:

> STEP 1. From its buffer of requests, the adversary selects one request of form $E(u, v, p)$ and delivers $p$ to $v$, and also selects one request of form $E(v, u, p')$ and delivers $p'$ to $u$. If there are no requests $(u, v, p)$ (resp. $(v, u, p')$), then the adversary sets $p$ (resp. $p'$) to $\bot$.

> STEP 2. Node $u$ (resp. $v$) sends new requests to the adversary of form $(u, v, p)$ (resp. $(v, u, p')$).

Note that the two above-mentioned actions take place *sequentially*, so that the requests queued to the adversary in Step 2 can depend on the packets received in Step 1, but requests formulated during Step 2 of some round $E(u, v)$ will not be delivered until edge $E(u, v)$ is honored again (at the earliest). Since nodes in the network only send/receive packets when they are at one end of an edge currently being honored, nodes will not do anything except when they are a part of an honored edge. Thus, in describing Slide+, we need only describe what an internal node $u$ will do when it is part of an honored edge $E(u, v)$. Recall that $C$ denotes the size of each node's memory[27], and for simplicity we will assume that $C/n \in \mathbb{N}$, and also for Slide+, we will require $C \geq 8n^2$.

**Slide+ Protocol Description.**
During honored edge $E(u, v)$, let $(v, u, (p', h))$ denote the message that $u$ receives from $v$ in Step 1 of the round (via the scheduling adversary). Also, $u$ has recorded the request $(u, v, (p, H))$ that it made during Step 2 of the previous round in which $E(u, v)$ was honored; note that $v$ will be receiving this message during Step 1 of the current round.

1. If $u$ is the Sender, then:
   (a) If $h < C$, then $u$ deletes packet $p$ from his input stream $\{p_1, p_2, \ldots\}$ (and ignores the received packet $p'$), and then proceeds to Step (c).
   (b) If $h \geq C$, then $u$ keeps $p_i$ (and ignores the received packet $p_j$), and proceeds to Step (c).
   (c) The Sender finds the next packet $p_i \in \{p_1, p_2, \ldots\}$ that has not been deleted and is not currently an outstanding request already sent to the adversary, and sends the request $(u, v, (p_i, C + \frac{C}{n} - 1))$ to the adversary. Also, $u$ will update the fact that the current message request sent to $v$ is $(u, v, (p_i, C + \frac{C}{n} - 1))$.

2. If $u$ is the Receiver, then $u$ sends the request $(u, v, (\bot, \frac{-C}{n}))$ to the adversary. Meanwhile, if $p' \neq \bot$, then $u$ stores/outputs $p'$ as a packet successfully received.

3. If $u$ is any internal node, then:
   (a) If $H \geq h + (C/n - 2n)$, then $u$ will ignore $p'$, delete $p$ and the "ghost packet associated to $p$" (see Step 3d below), and slide down any packets/ghost packets to fill any gaps created. Also, $u$ will update his height $H = H - 1$, and proceed to Step 3d below.
   (b) If $H \leq h - (C/n - 2n)$, then $u$ will keep $p$, and also store $p'$ in the stack location that $u$ had been storing the "ghost packet" for $p$ (see Step 3d below), deleting the ghost packet in the process. Also, $u$ will update his height $H = H + 1$, and proceed to Step 3d below.
   (c) If $|H - h| < C/n - 2n$, then $u$ will ignore packet $p'$ and keep $p$, but delete the "ghost packet" associated to $p$, and then proceed to Step 3d.

---
[27]For simplicity, we assume that all nodes have the same memory bound, although our argument can be readily extended to handle the more general case.

(d) Node $u$ will search its stack for the highest packet $p''$ (not including ghost packets) that it has not already committed in an outstanding request to the adversary. It then sends the request $(u, v, (p'', H))$ to the adversary. Additionally, $u$ will create a "ghost packet associated to the packet/request $p'''$" that it has just sent the adversary. This "ghost packet" will assume the first un-filled spot in $u$'s memory stack. Finally, $u$ will update the fact that the current message request sent to $v$ is $(u, v, (p'', H))$.

In the following section, we will prove that the above routing rules are compatible with memory requirements (e.g. that Steps 3b and 3d do not require a node to store more than $C$ (ghost) packets), as well as prove that Slide+ enjoys competitive ratio $1/n$.

## C.2 Analysis of Slide+

Before providing the full details of the proof that Slide+ enjoys competitive ratio $1/n$, we will provide a brief high-level description of how the proof works. First, notice that the main technical challenge in moving from the semi-asynchronous model of Section 4 to the fully asynchronous model is that nodes can no longer make routing decisions based on *current* information. Indeed, the current state of a node may change drastically from the time it makes a request in Step 2 of some round $E(u, v)$ and the time the request is finally sent by the adversary in Step 1 of the next round in which $E(u, v)$ is honored. Since the Slide protocol uses the current height of a node to make routing decisions, the fact that the height of a node may change substantially between the time a packet request is made and the time the receiving node receives the packet is an issue that must be resolved.

The above described protocol handles this issue by allotting "ghost packets" in Step 3d (this will ensure there is always room to store a packet sent from an honest neighbor), as well as having nodes make routing decisions based on *old* height considerations. In particular, Steps 1-3 above dictate what $u$ should do based on the height that $u$ and $v$ had during *the last time $E(u, v)$ was honored*. Therefore, although this information may have become outdated since the last time $u$ and $v$ communicated with each other, at least the decisions will be made *consistently*, both in the sense that the heights being compared are *synchronized* (i.e. they are from the same time as each other, although possible now out-dated), and in the sense that the nodes will know what the other will do (assuming both are honest) in terms of whether or not it will keep the packet just sent/received. This last fact is crucial to prevent packet deletion and duplication from occurring (at least amongst honest nodes).

The proof will follow the main structure of the proof provided for the semi-asynchronous Slide protocol, with one additional category to account for packet transferring decisions that were based on significantly outdated height information.

**Theorem C.1.** *The Slide+ protocol achieves competitive ratio $1/n$ in any distributed, asynchronous, bounded memory network with dynamic topology (and no minimal connectivity assumptions). More specifically, for any adversary/off-line protocol pair $(\mathcal{A}, \mathcal{P}')$, if $\mathcal{P}$ denotes the Slide+ protocol, $C$ denotes the capacity (memory bound) of each node, and $Z_x^{\mathcal{P}}$ (resp. $Z_x^{\mathcal{P}'}$) denotes the number of packets received by protocol $\mathcal{P}$ (resp. $\mathcal{P}'$) as of round $x$, then for all rounds $x$:*

$$Z_x^{\mathcal{P}'} \leq 8n Z^{\mathcal{P}} + 8n^2 C \tag{74}$$

*Proof.* Fix any adversary/off-line protocol pair $(\mathcal{A}, \mathcal{P}')$, and let $\mathcal{P}$ denote the Slide+ protocol and $Z_x^{\mathcal{P}}$ and $Z_x^{\mathcal{P}'}$ as in the statement of the theorem. Motivated by the proof in the semi-asynchronous setting, we imagine a virtual world in which the two protocols are run simultaneously in the same network. We split $Z_x^{\mathcal{P}'}$ into the following three subsets (we will henceforth suppress the index referencing the round $x$):

1. $Z_1^{\mathcal{P}'}$ consists of packets $p' \in Z^{\mathcal{P}'}$ for which there exists at least one round $E(u, v)$ such that both $p'$ was transferred by $\mathcal{P}'$ *and* some packet $p$ was transferred by $\mathcal{P}$.[28]

---

[28] Note that we make no condition that the two packets traveled in the same direction.

2. $Z_2^{\mathcal{P}'}$ consists of packets $p' \in Z^{\mathcal{P}'}$ that were *never* transferred alongside a packet in $\mathcal{P}$ as in 1 above, and such that *every time $p'$* was transferred between two nodes $u$ and $v$ during a round $E(u,v)$, the heights $H$ and $h$ that were used by $u$ and $v$ in determining whether to store/delete the packets delivered by the adversary during Step 1 of $E(u,v)$ (see protocol description above) were each within $n$ of the current heights of $u$ and $v$.

3. $Z_3^{\mathcal{P}'} = Z^{\mathcal{P}'} \setminus (Z_1^{\mathcal{P}'} \cup Z_2^{\mathcal{P}'})$.

Clearly, $|Z^{\mathcal{P}'}| = |Z_1^{\mathcal{P}'}| + |Z_2^{\mathcal{P}'}| + |Z_3^{\mathcal{P}'}|$, and hence the Theorem follows from Lemmas C.3, C.4, and C.5 below. ∎

We will need the following trivial observation, which follows immediately from the description of the Slide+ protocol in Section C.1.

**Observation 3.** *At all times, an internal node $u$ has at most $n$ ghost packets and at most $n$ outstanding requests (one for each of its edges $v$).*

*Proof.* Rules 1(c) and 3(d) only allow a node to submit a single request for each round the node is part of an honored edge, and this request is then delivered by the adversary in Step 1 of the next round in which the edge is honored. Also, Rules 3(a-c) guarantee that the ghost packet corresponding to the current honored edge will be deleted before another one is created in Rule 3(d). ∎

In order to bound $|Z_1^{\mathcal{P}'}|$, we will need to bound the number of times any packet $p$ can be transferred by the Slide+ protocol. In the asynchronous Slide protocol of Section 4, we showed that any packet $p$ could be transferred at most $2n$ times, as during every packet transfer in Slide, the packet must drop in height by at least $C/n-1$. At first glance, it might seem that we cannot make the same argument in the fully asynchronous setting since the Slide+ protocol is making routing decisions based on (potentially) outdated height information. However, the introduction of "ghost packets" will allow us to retain this quality. Indeed, the purpose of utilizing ghost packets is to anticipate future packet transfers and reserve spots in a node's memory stack at the appropriate height, allowing us to argue that even if nodes nodes are using out-dated height information, packets will still "flow downhill" from Sender to Receiver. This is captured in the following lemma.

**Lemma C.2.** *Let $Y_x^{\mathcal{P}}$ denote the the set of packets inserted by $\mathcal{P}$ as of round $x$. Also let $T_x^{\mathcal{P}}$ denote the set of packet transfers that have occurred in $\mathcal{P}$ as of round $x$. Then any packet in the Slide+ protocol is transferred at most $2n$ times.[29] In particular, $|T_x^{\mathcal{P}}| \leq 2n|Y_x^{\mathcal{P}}| \leq 2n(|Z_x^{\mathcal{P}}| + nC)$.*

*Proof.* We show that anytime a packet is transferred in the Slide+ protocol, the packet's height in the new buffer is necessarily at least $C/n - 4n$ *lower* than its height in the old buffer. Since packets only move within buffers when they are received or sent (or when they slide *down* as in 3(a)), and since[30] $2n(C/n - 4n) > C$, the lemma will follow. Fix a packet $p$, and consider a round $x = E(u,v)$ in which $p$ is transferred from $u$ to $v$. In particular, it must have been that the *previous* round $x' < x$ in which $E(u,v)$ was honored, $u$ sent some request of form $(u,v,(p,H))$ to the adversary in Step 2. Notice that when $u$ selected $p$ to form a part of its request as in 3(d), since $u$ had height $H$ and $u$ has at most $n-1$ packets already committed as an outstanding request (Observation 3), $p$ must have height at least $H - n$ in $u$'s buffer. Meanwhile, let $(v,u,(p',h))$ denote the request that $v$ sent to the adversary in Step 2 of round $x'$. Notice that in 3(d), $v$ reserved a position in its buffer (the "ghost packet"), into which $p$ will be inserted when it is received in round $x$. Since the ghost packet is assigned the topmost unoccupied (by packet or ghost packet) position in $v$'s buffer, we have that $p$ will have height no bigger than $h + n$. Therefore, $p$ will drop in height by at least $(H-n) - (h+n) = H - h - 2n$ when it is transferred from $u$ to $v$. Since the criterion for accepting a new packet (see 3(d)) demands that $H - h \geq C/n - 2n$, we have that $p$ will necessarily drop in height by at least $C/n - 4n$ when it is transferred. ∎

---
[29]This matches the bound for the semi-asynchronous Slide protocol of Section 4.
[30]For Slide+, we have demanded that $C > 8n^2$.

Notice that Lemma C.2 is valid *regardless* of how long a request $(u, v, (p, H))$ has been queued in the adversary's buffer, and also of how $u$ and $v$'s stacks may have changed in the meantime. We are now ready to state and prove the first requisite bound:

**Lemma C.3.** $|Z_1^{\mathcal{P}'}| \leq 2n|Z^{\mathcal{P}}| + 2n^2 C$

*Proof.* By definition, $|Z_1^{\mathcal{P}'}| \leq |T^{\mathcal{P}}|$, and the latter is bounded by $2n|Z^{\mathcal{P}}| + 2n^2 C$ by Lemma C.2. ∎

**Lemma C.4.** $|Z_2^{\mathcal{P}'}| \leq 2n|Z^{\mathcal{P}}| + 2n^2 C$

*Proof.* This bound follows the same reasoning as the proof of Lemma B.18. Suppose that packet $p' \in Z_2^{\mathcal{P}'}$ is transferred by $\mathcal{P}'$ from $u$ to $v$ in round $x$. By definition of $Z_2^{\mathcal{P}'}$, Slide+ did not transfer a packet, and thus (with the notation as in Rule 3(d) for Slide+) $|H - h| < C/n - 2n$. Also by definition of $Z_2^{\mathcal{P}'}$, we have that $v$'s height in round $x$ is within $n$ of $h$, and $u$'s height in round $x$ is within $n$ of $H$. Consequently, $u$'s height in round $x$ must be within $C/n$ of $v$'s height. Then if we define families the same way as in the proof for the semi-synchronous Slide protocol (see Section B), by Lemma B.12, $u$ and $v$ must be in the same family at the start of $x$. Indeed, all the lemmas and proofs of Section B will remain valid[31], and hence Lemma B.18, which states that $|Z_2^{\mathcal{P}'}| \leq 2n|Z^{\mathcal{P}}| + 2n^2 C$, remains valid. ∎

**Lemma C.5.** $|Z_3^{\mathcal{P}'}| \leq 4n|Z^{\mathcal{P}}| + 4n^2 C$

*Proof.* Fix a packet $p' \in Z_3^{\mathcal{P}'}$. By definition of $Z_3^{\mathcal{P}'}$, there exists some round $x_{p'} = E(u, v)$ in which $p'$ was transferred from $u$ to $v$, where either $u$'s height or $v$'s height has changed by at least $n$ since the previous round $x'_{p'} < x$ in which $E(u, v)$ was honored. Let $\mathcal{S}_{p'} \subseteq T^{\mathcal{P}}$ denote $n$ of these packet transfers, where each packet transfer in $\mathcal{S}_{p'}$ corresponds to a packet sent (or received) by $u$ (or $v$), and took place between $x'_{p'}$ and $x_{p'}$.

> **Observation.** For any packet transfer in Slide+, there are at most $2n$ packets $p' \in Z_3^{\mathcal{P}'}$ for which the packet transfer appears in $\mathcal{S}_{p'}$.
>
> *Proof.* Consider any round $x' = E(u, v)$ in which a packet is transferred from $u$ to $v$ by Slide+, and refer to this specific packet transfer as $T_{x'}$. Then for each edge of $u$ and each edge of $v$ and for any $p' \in Z_3^{\mathcal{P}'}$, there can be at most one round $x_{p'} > x'$ for which $T_{x'} \in \mathcal{S}_{p'}$. After all, once a given edge of $u$ or $v$, say for example $E(u, w)$, transfers a packet $p' \in Z_3^{\mathcal{P}'}$ in round $x_{p'} > x'$, the heights of both $u$ and $w$ are updated, and there can never be another $p'' \in Z_3^{\mathcal{P}'}$ and later round $x_{p''} > x_{p'}$ such that $x_{p''} = E(u, w)$ and $T_{x'} \in \mathcal{S}_{p''}$. Therefore, $T_{x'}$ can appear in at most $2n$ sets of form $\mathcal{S}_{p'}$.

Since $|\mathcal{S}_{p'}| = n$ for each $p' \in Z_3^{\mathcal{P}'}$, we have that:

$$\sum_{p' \in Z_3^{\mathcal{P}'}} |\mathcal{S}_{p'}| = n|Z_3^{\mathcal{P}'}| \tag{75}$$

Now since for any given packet transfer $T_x \in T^{\mathcal{P}}$ there can be at most $2n$ different values of $p' \in Z_3^{\mathcal{P}'}$ such that $T_x \in \mathcal{S}_{p'}$, we have that:

$$\left| \bigcup_{p' \in Z_3^{\mathcal{P}'}} \mathcal{S}_{p'} \right| \geq \frac{n|Z_3^{\mathcal{P}'}|}{2n} \tag{76}$$

But $\cup_{p' \in Z_3^{\mathcal{P}'}} \mathcal{S}_{p'} \subseteq T^{\mathcal{P}}$, so:

$$|T^{\mathcal{P}}| \geq |\cup_{p' \in Z_3^{\mathcal{P}'}} \mathcal{S}_{p'}| \geq \frac{|Z_3^{\mathcal{P}'}|}{2} \tag{77}$$

In particular, $|Z_3^{\mathcal{P}'}| \leq 2|T^{\mathcal{P}}| \leq 4nZ^{\mathcal{P}} + 4n^2 C$, where the second inequality is Lemma C.2. ∎

---

[31] The only necessary modification is to consider the present definition of $Z_2^{\mathcal{P}'}$ instead of the one used in Section B